\documentclass[letterpaper]{article}

\usepackage[T1]{fontenc}

\usepackage{geometry}
\usepackage{setspace}

\usepackage{subcaption}  % Place this in the preamble
\usepackage{lineno}
\usepackage[percent]{overpic} % in preamble
\usepackage{comment}

\usepackage{titlesec}
\titleclass{\subsubsubsection}{straight}[\subsubsection]
\newcounter{subsubsubsection}[subsubsection]
\renewcommand\thesubsubsubsection{\thesubsubsection.\arabic{subsubsubsection}}
\titleformat{\subsubsubsection}
  {\normalfont\normalsize\bfseries}{\thesubsubsubsection}{1em}{}
\titlespacing*{\subsubsubsection}{0pt}{3.25ex plus 1ex minus .2ex}{1.5ex plus .2ex}

\makeatletter
\def\toclevel@subsubsubsection{4}
\def\l@subsubsubsection{\@dottedtocline{4}{7em}{4em}}
\makeatother

\usepackage{xcolor}      % MUST come before definecolor
\usepackage{sectsty}

\definecolor{myblue}{RGB}{0,102,204}
\definecolor{myred}{RGB}{200,50,50}

\sectionfont{\color{myred}}          % sections
\subsectionfont{\color{black}}       % subsections
\subsubsectionfont{\color{myblue}}   % subsubsections

\usepackage[style = chem-acs]{biblatex} 
\usepackage{graphicx}
\usepackage{float}
\newfloat{scheme}{htbp}{los}
\floatname{scheme}{Scheme}
\floatname{chart}{Chart}
\newfloat{graph}{htbp}{loh}

\usepackage{chemformula} % Formulas using \ch{}
\usepackage[version = 4]{mhchem} % Formulas using \ce{}

\usepackage{authblk} 
\author[1,2]{ Mohammad Tahsin Alam\thanks{These authors contributed equally to this work}} 
\author[1]{Zafrin Jahan Nikita\thanks{These authors contributed equally to this work}} \author[3]{Ying Yin Tsui} 
\author[1]{Md Zahurul Islam} 
\affil[1]{Department of Electrical and Electronic Engineering, Bangladesh University of Engineering and Technology, Dhaka 1205, Bangladesh} 
\affil[2]{Department of Computer Science and Engineering, BRAC University, Dhaka, Bangladesh} 
\affil[3]{Department of Electrical and Computer Engineering, University of Alberta, Edmonton, AB T6G 2H5 Canada}

\title{Tamm Plasmon–Enhanced Widely Tunable Near-Infrared Nanolaser with Superior Efficiency and Output Power}
\vspace{-0.5cm}

\date{*Email: mdzahurulislam@eee.buet.ac.bd}

\usepackage[colorlinks=true,
            linkcolor=black,
            citecolor=black,
            urlcolor=black]{hyperref}

\begin{document}

\maketitle

\vspace{-0.7cm}
\begin{abstract}
  Plasmonic resonances enable strong electromagnetic field confinement and have been widely exploited in plasmonic nanolasers, particularly through surface plasmon polaritons and localized surface plasmons. However, their performance is often limited by bidirectional output coupling and multimode far-field emission, primarily due to higher-order diffraction arising from these modes. In this work, we utilize the Tamm plasmon resonance to realize lasing in the NIR region with wide tunability. The optical Tamm states are excited at the metal-DBR interface by an incident pump pulse and their emission intensity is significantly enhanced via extraordinary optical transmission through a metallic nanohole array. The subwavelength periodicity of the nanohole array restricts the emission to the zeroth order, resulting in a highly directional far-field pattern with a FWHM of approximately $0.631^\circ$.To further improve performance, a second DBR is incorporated beneath the pump side, which substantially suppresses backward emission around the lasing wavelength and enhances forward lasing intensity by around $1.3 \times 10^{4}$ times, thus increasing the integrated emission power. The combination of Tamm plasmon excitation and dual-DBR feedback significantly improves the cavity's optical response and overall lasing efficiency. Additionally, we have demonstrated lasing at 870 nm with a reduced pump threshold of $2.8 \times 10^{7} \text{V/m}$ (energy of $0.0031 \text{mJ/cm}^{2}$). Moreover, a broad tunability in lasing wavelength, spanning from 850 nm to 944.5 nm is achieved. These results demonstrate a cost-effective and versatile strategy for plasmonic nanolasers with enhanced output power, low reflection-side loss, wide tunability and strong integration potential for on-chip photonic and quantum technologies.
\end{abstract}

\vspace{0.1cm}


%%%%%%%%%%%%%%%%%%%%%%%%%%  body  %%%%%%%%%%%%%%%%%%%%%%%%%%

\vspace{-0.35cm}

\section{1. Introduction}

\vspace{-0.2cm}  % reduces space by 0.5 cm

Integrated photonics increasingly uses nanoscale plasmonic lasers as compact on-chip optical sources.\cite{shahid_merged_2022} Unlike conventional diffraction-limited lasers,\cite{ning2010nanolasers} plasmonic nanolasers excite surface plasmon polaritons (SPPs) and localized surface plasmons (LSPs),\cite{ahamed_wavelength_2024} confining light in nanogaps or nanoholes for strong field enhancement and efficient gain interaction. Despite microscale cavities, lasing occurs in sub-diffraction volumes, producing coherent emission from extremely small optical modes. These devices are promising for biomedical sensing, imaging, spectroscopy, nanolithography, high-performance on-chip links, and photonic-electronic integration.

 SPP modes enable nanoscale light confinement but suffer from intrinsic ohmic and surface-scattering losses, often requiring cryogenic operation, high-gain media, and wavevector-matching schemes.\cite{ahmed_efficient_2018, ding_metallic_2012, berini_surface_2012, Bouhelier_2005, maier_plasmonics_2005, kar_tamm_2023} Far-field emission is typically highly divergent due to mismatch between plasmonic and free-space wavevectors. Plasmonic crystal-based nanolasers using periodic metallic arrays or nanoholes achieve directional lasing via band-edge lattice plasmon and SPP-Bloch/EOT modes, producing narrow divergence (1-3°) and strong coherence.\cite{zhou_lasing_2013, symonds_confined_2013, zhang_lasing_2016, vanbeijnum_surface_2013, vanbeijnum_loss_2014, meng_highly_2014, ebbesen_extraordinary_1998, meng_wavelength_2013} However, emission often spreads over multiple diffracted modes from top and bottom surfaces; unidirectional emission can be realized with thick-substrate 2D plasmonic crystals, albeit with pump-light separation and ohmic-loss limitations.\cite{yang_unidirectional_2015}

 To overcome these issues, several Tamm plasmon-based lasers have already been investigated. When uniform subwavelength hole arrays on thin metal films are placed on a DBR, such films support Tamm plasmon (TP) states~\cite{melentiev_single_2011, treshin_optical_2013, kavokin_lossless_2005, kaliteevski_tamm_2007} with the nanohole array enhancing light transmission,\cite{ebbesen_extraordinary_1998} with EOT influenced by hole shape, size, periodicity, metal thickness, and material.\cite{koerkamp_strong_2004, vander_molen_role_2005, genet_light_2007, giannattasio_transmission_2004} Semiconductor-based confined TP lasers have shown lasing near 857 nm using InGaAs QWs~\cite{symonds_confined_2013} and ZnO UV lasers at 372 nm via exciton-TPP coupling~\cite{xu_tamm_2022}. Simulation studies demonstrated single-mode NIR lasers with high directionality (<1$^\circ$) using EOT and OTS feedback~\cite{ahmed_efficient_2018}, with tunable wavelengths via gain/thickness and pump angle adjustments. Dual-mode NIR lasers with merged NHAs achieved <0.35° divergence and tunable emission~\cite{shahid_merged_2022}, while Shahid and Talukder highlighted that hole arrangement, from ordered to random, strongly impacts laser modes and performance~\cite{shahid_beyond_2025}. However, alternative metastructures other than square-shaped holes and strategies to suppress backward emission for low-threshold, widely tunable, high-output single-mode lasers remain unexplored.

 We propose a cost-effective, high-power plasmonic nanolaser (PNL) operable at room temperature in the NIR range with wide tunability. Using EOT and OTS, transmission is enhanced through a periodic array of octagonal nanoholes. The design employs top and bottom DBRs: the top DBR couples to an IR-140-doped PU gain medium with a terminating layer, while the gain layer is separated from a 100 nm metal coating fully covered with alumina and etched with octagonal holes. The top 1DPC layers enable stimulated emission to couple to OTS at the metal-DBR interface, further amplifying EOT via NHAs. Initially, a single-mode structure with a top 1DPC and silver-coated metal layer is analyzed, highlighting its limitations. A dual-DBR model is then proposed to boost emission and minimize losses. Lasing wavelength can be tuned via various structural parameters.

\vspace{-0.4cm}

\section{2. Proposed Structure and Methodology}

\vspace{-0.08cm}

%\subsection{Proposed Structure}
\label{Design_params}
\vspace{-0.12cm}

\begin{figure}[htbp]
    \centering
    % Left figure
    \begin{subfigure}[b]{0.48\linewidth}
        \centering
        \includegraphics[width=\linewidth, trim=160 10 150 70]{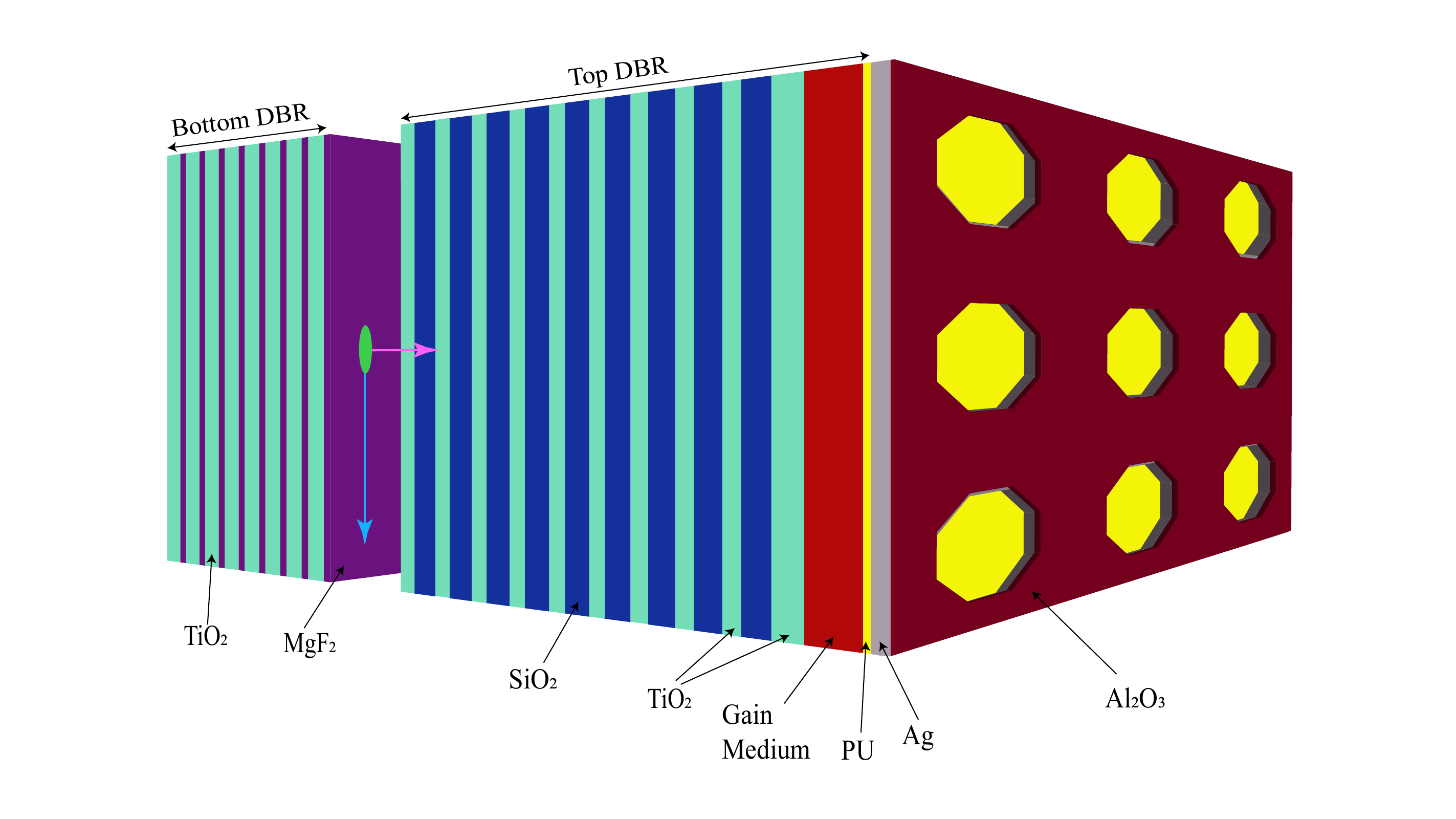}
        \captionsetup{justification=centering}
        \caption{3D illustration of the proposed PNL. }
        \label{perspective}
    \end{subfigure}
    \hfill
    % Right figure
    \begin{subfigure}[b]{0.48\linewidth}
        \centering
        \includegraphics[width=1\linewidth, trim=50 100 50 150]{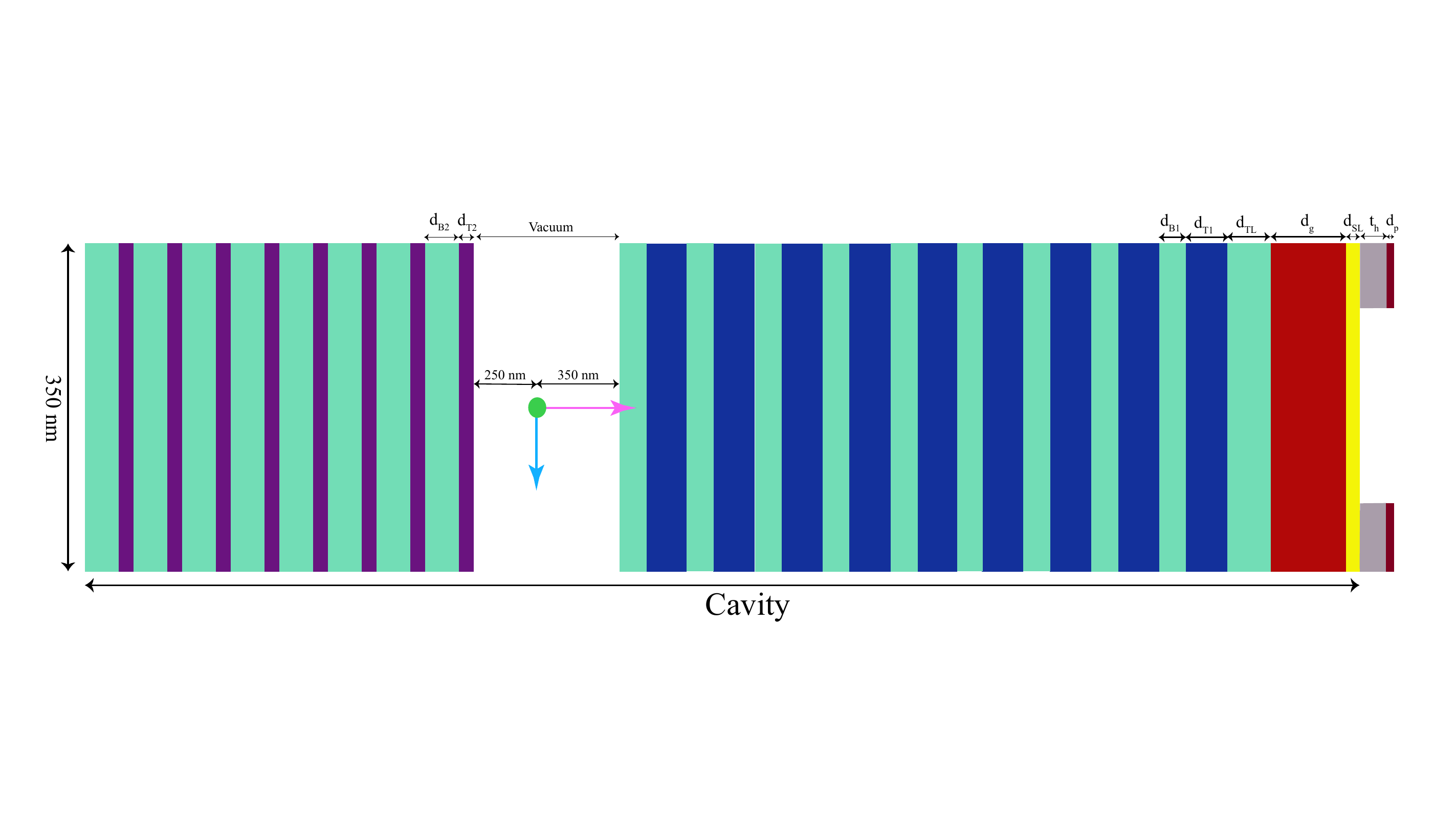}
        \captionsetup{justification=centering}
        \caption{Cross-sectional view of a unit cell of the proposed model.}
        \label{cross_section}
    \end{subfigure}

    \caption{Schematic of the proposed Dual-DBR Laser model.\vspace{-0.2cm}}
    \label{Overall_model}
    \end{figure}

Figure~\ref{perspective} illustrates a 3D schematic of the proposed nanolaser, whereas Figure~\ref{cross_section} presents a cross-sectional view of a single unit cell along the x-z plane. The overall design comprises two main components: the Tamm plasmon generation section and a bottom reflector, with the pump source positioned between them.

The Tamm plasmon generation side consists of a metastructure consisting of a 100~nm thick Ag layer ($t_\text{h}$) coated with a 1~nm alumina layer ($d_\text{p}$), patterned with a periodic octagonal nanohole array of 102.5~nm diameter ($d_\text{h}$), 101~nm depth, and 350~nm periodicity. A 20~nm polyurethane (PU) separation layer ($d_\text{SL}$) and a 310~nm($d_\text{g}$) IR-140-doped PU gain layer are placed above it. This stack rests atop a \textit{9-layer} 1D photonic crystal comprising alternating SiO\textsubscript{2} (top) and TiO\textsubscript{2} (bottom) layers of thicknesses $d_\text{T1}$ = 170~nm and $d_\text{B1}$ = 111~nm, respectively, terminated with a TiO\textsubscript{2} layer of $d_\text{TL}$ = 186~nm. The Ag layer is coated with alumina layer to prevent oxidation and enhance thermal and chemical stability. In Tamm plasmon lasers, direct gain-metal contact increases ohmic loss due to strong field penetration. As suggested by Zabir et al.\cite{ahmed_design_2016}, a thin dielectric spacer (d\textsubscript{SL}) shifts the optical mode away from the metal, reducing absorption while maintaining Tamm plasmon coupling.

The pump source is located 350~nm below the top structure. The bottom DBR is positioned 250~nm below the pump source, following the tuning of the top structure for Tamm plasmon resonance at 870~nm. It comprises \textit{8} alternating layers of MgF\textsubscript{2} (top) and TiO\textsubscript{2} (bottom) with thicknesses of $d_\text{T2}$ = 60~nm and $d_\text{B2}$ = 140~nm, respectively. The region between the two DBRs is filled with vacuum to allow pump incidence. The gain medium parameters are adopted from Zabir et al.\cite{ahmed_efficient_2018} Optimization of all layers is discussed in section \ref{Optimization}.

We numerically solved the complete device in three dimensions by directly solving Maxwell’s electromagnetic equations using Lumerical FDTD Solutions, a full-vector electromagnetic simulation tool. In the simulation, we assumed our proposed nanolaser to be infinitely periodic along the x- and y-directions. Since our structure is antisymmetric along x and symmetric along y with respect to the electric field of the applied pump source, we set antisymmetric and symmetric boundary conditions along the x- and y-axes, respectively, which also account for the periodicity. Along the z-direction, a Perfectly Matched Layer (PML) boundary condition was applied. We performed the FDTD simulation only for a single unit cell with a periodicity of 350 nm in both the x- and y-directions to calculate the overall response of the entire structure, taking advantage of its periodicity. We used a plane wave source as the optical pumping mechanism, centered at 800 nm, with a pulse length of 40 fs and an offset of 80 fs. The emission behavior observed via DFT monitors was evaluated by computing the power flux through the field monitor using the normal components of the Poynting vector.  

The electromagnetic solver incorporates a multi-coefficient dispersion model to represent the optical response of silver in the near-infrared range. The material parameters for silver are adopted from Johnson and Christy.\cite{johnson1972optical} The dielectric materials used in the proposed design are considered non-dispersive within the NIR range, with refractive indices taken from~\cite{melentiev_single_2011},\cite{ahmed_efficient_2018}.The base material for the gain medium is PU, since IR-140 dyes are embedded in PU. In dye molecules, absorption and fluorescence take place across four singlet energy levels, which allows dye lasers to be described using a four-level model.\cite{nair_dye_1982} In this work, we represent IR-140 with a semi-quantum mechanical framework that treats it as a four-level two-electron system.\cite{chang_fdtd_2004} Here, the dye is modeled quantum mechanically, while the electromagnetic field is described in classical terms. The temperature was kept at standard room temperature, 300~K.

 \begin{comment}

\begin{table}[htbp]
    \centering
    \caption{Refractive indices of dielectric layers.} 
    \label{refractive_indices}
    \begin{tabular}{@{}c c@{}}
        \hline
        \textbf{Material} & \textbf{Refractive Index} \\ \hline
        TiO\textsubscript{2} & 2.23 \\
        MgF\textsubscript{2} & 1.38 \\
        PU (Polyurethane)    & 1.51 \\
        SiO\textsubscript{2} & 1.44 \\ \hline
    \end{tabular}
\end{table}

\end{comment}

\vspace{-0.3cm}

\section{3. Results and Discussions}
\vspace{-0.2cm}

\subsection{\textcolor{myred}{\textbf{3.1. Toward building the optimized PNL model}}}
\label{Optimization}
\vspace{-0.14cm}

In this section, we optimize and construct the complete laser model step by step to achieve lasing at 870 nm. The analysis emphasizes the choice of hole geometry, adjustment of metal width, determination of DBR layer thicknesses, and the role of the bottom reflector before finalizing the design.

\vspace{-0.3cm}

\subsubsection{3.1.1    Silver (coated) nanohole demonstrating EOT}

    \vspace{-0.15cm}

\begin{figure}[htbp]
    \centering

    \begin{subfigure}[b]{0.32\textwidth}
        \centering
        \includegraphics[width=.8\textwidth, trim=410 80 380 50]{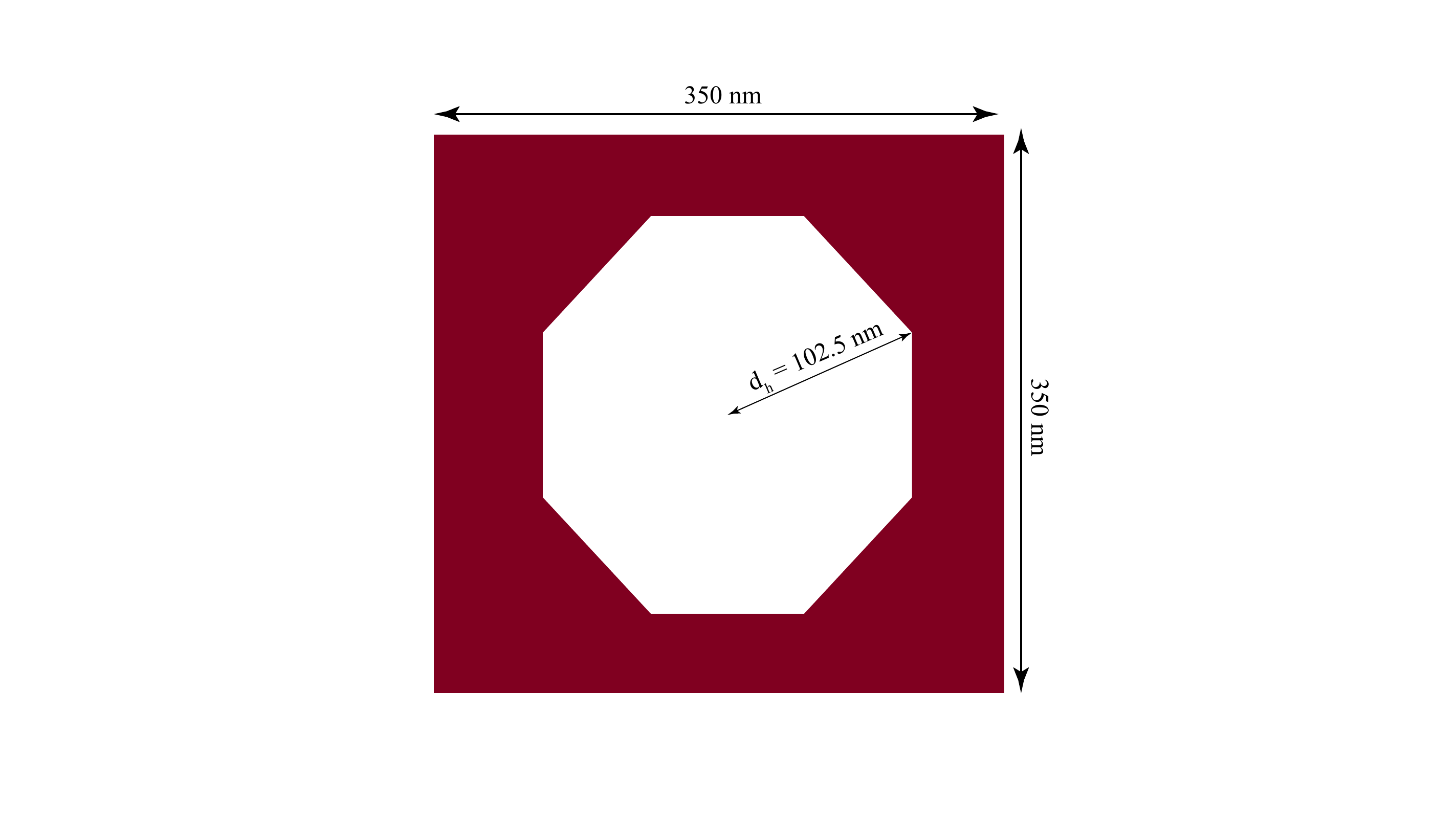}
        \captionsetup{justification=centering}
        \caption{Dimension of the nanocavity.}
        \label{hole_dim}
    \end{subfigure}%
    \hspace{0.01\textwidth}%
    \begin{subfigure}[b]{0.32\textwidth}
        \centering
        \includegraphics[width=\textwidth, trim=50 40 90 90]{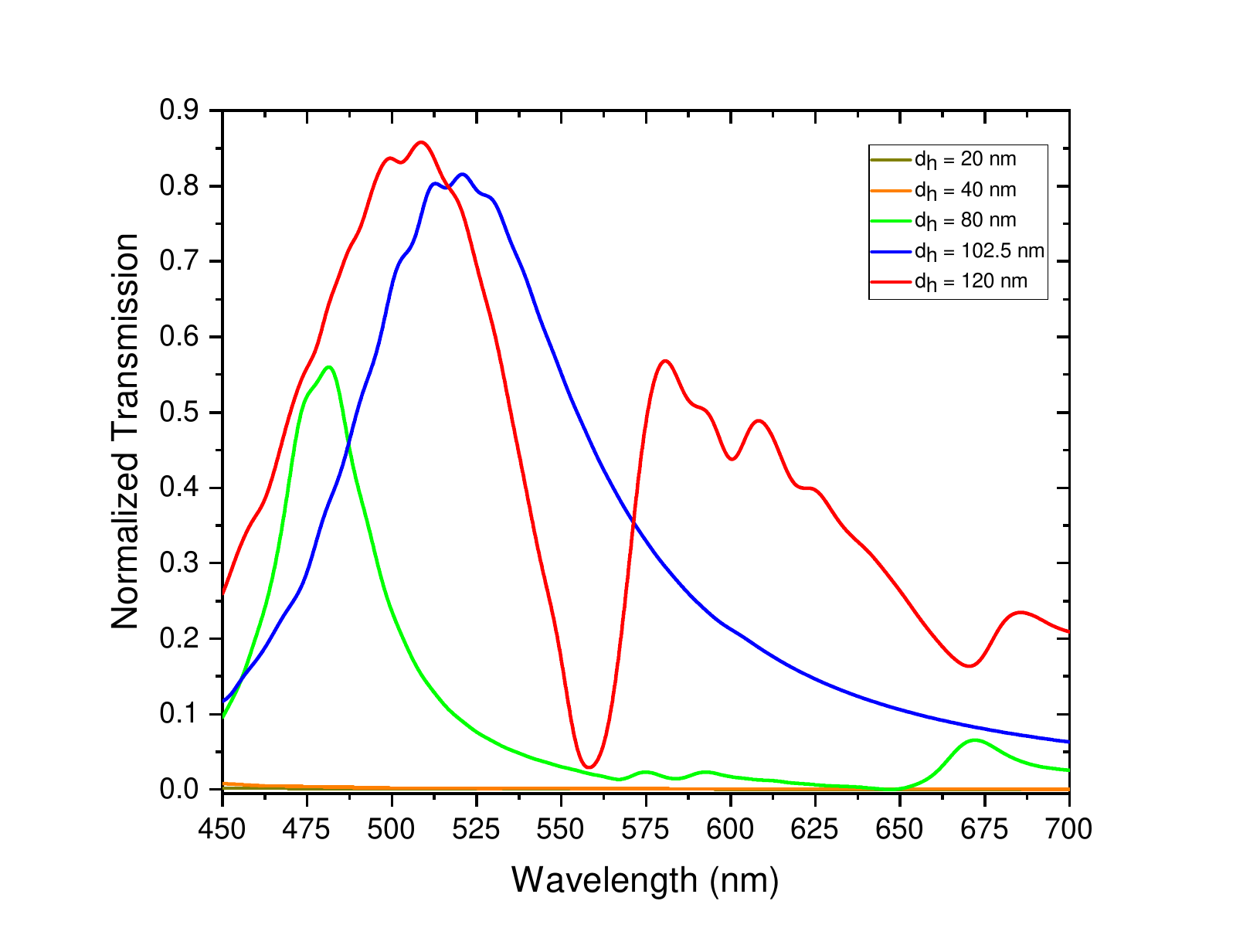}
        \captionsetup{justification=centering}
        \caption{Transmission for varying  $d_h$.}
        \label{holesize_vary_EOT}
    \end{subfigure}%
    \hspace{0.01\textwidth}%
    \begin{subfigure}[b]{0.32\textwidth}
        \centering
        \includegraphics[width=\textwidth, trim=50 40 90 90]{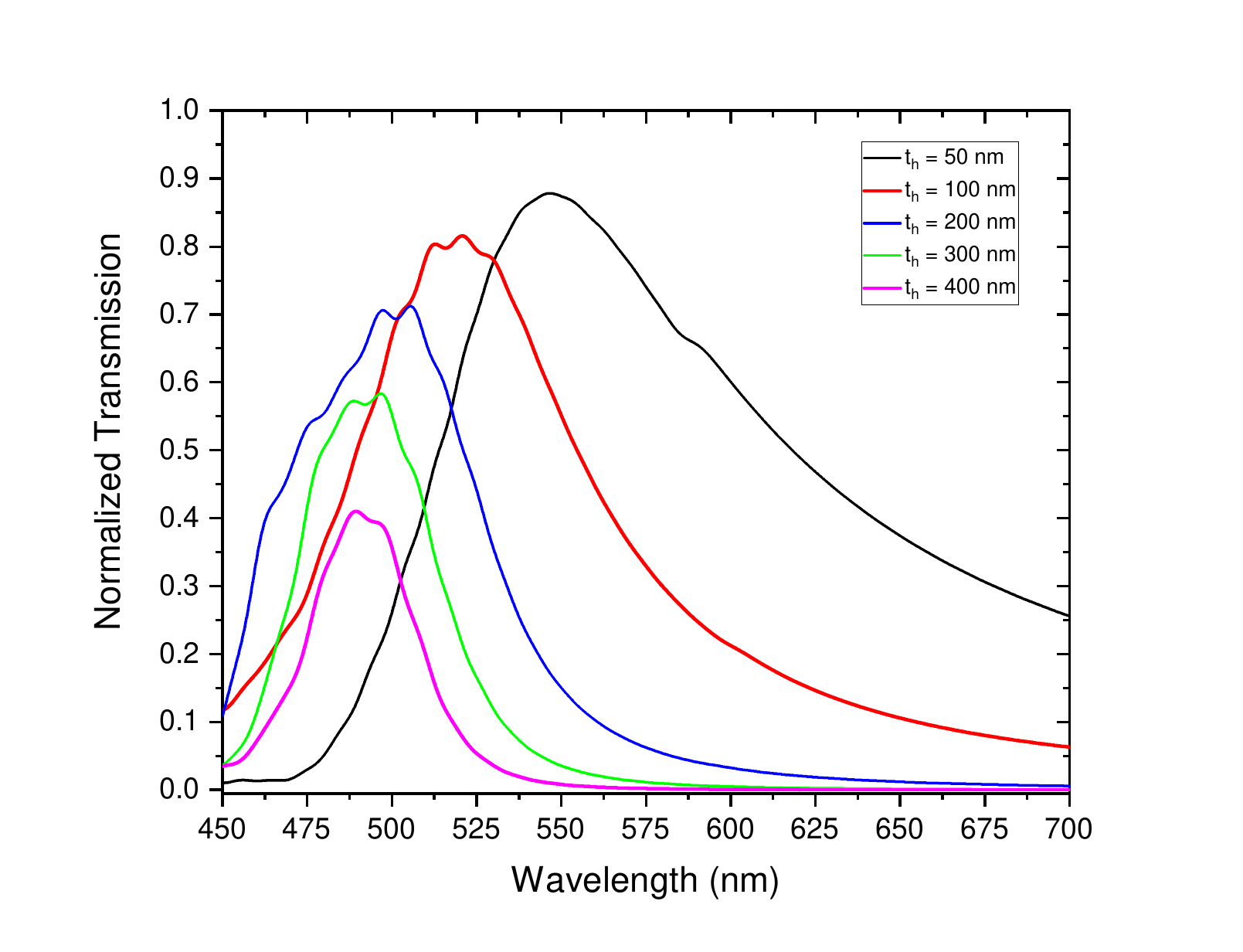}
        \captionsetup{justification=centering}
        \caption{Transmission for varying $t_h$.}
        \label{th_vary_EOT}
    \end{subfigure}

    \caption{Selection of $t_h$ and $d_h$ based on EOT via a single periodic nanohole.\vspace{-0.4cm}}
    \label{EOT_all}
\end{figure}

    A typical EOT spectrum shows multiple peaks and troughs with intensities exceeding classical aperture predictions for subwavelength holes. This study focuses on subwavelength-periodicity arrays, where EOT is dominated by localized resonances of individual holes, enabling coupling into a single diffracted mode. Larger periods may enhance collective resonances but introduce higher-order diffraction, reducing lasing efficiency. In this section, we observed transmission via a single nanohole etched into coated silver layer as depicted in figure \ref{hole_dim}. For hole size $d_h = 102.5$ nm, a pronounced narrow peak is observed (Figure~\ref{holesize_vary_EOT}), and a metal thickness of 100 nm is chosen based on Figure~\ref{th_vary_EOT}, which shows the strongest response for $t_h = 100$ nm. Increasing $d_h$ or decreasing $t_h$ broadens the spectrums, potentially generating multi-order lasing modes as seen in figure \ref{holesize_vary_EOT} and \ref{th_vary_EOT}. Key mechanisms behind the realization of EOT include coupling of incident light to surface plasmon polaritons (SPPs) via the hole lattice, excitation of localized resonances in each hole, and constructive interference from coherent scattering and re-radiation across the periodic array. \cite{ebbesen_extraordinary_1998, genet_light_2007, van_der_molen_role_2005}

\vspace{-0.2cm}
\subsubsection{3.1.2.   Response of plasmonic cavity in absence of OTS}

\vspace{-0.2cm}

\begin{figure}[htbp]
    \centering
    % Left subfigure (PDF with trim)
    \begin{subfigure}[b]{0.48\linewidth}
        \centering
        \includegraphics[width=.8\linewidth, trim=170 5 170 5]{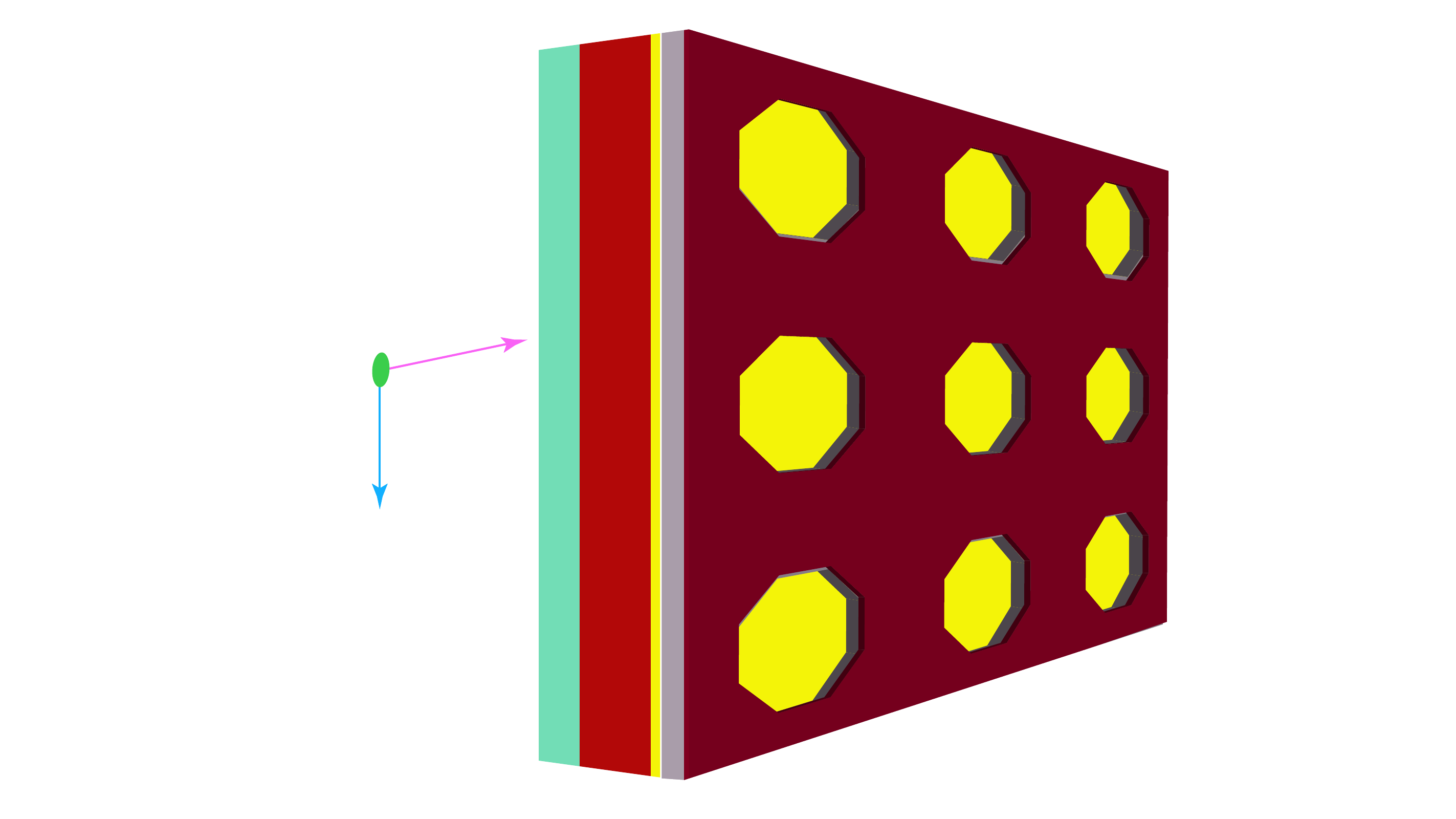}
        \captionsetup{justification=centering}
        \caption{Plasmonic nanolaser with perforated silver layer\\ followed by a gain medium and a terminating layer.}
        \label{EOT_based_cavity}
    \end{subfigure}
    \hfill
    % Right subfigure (JPEG, no trim)
    \begin{subfigure}[b]{0.48\linewidth}
    \centering
    \includegraphics[width=.74\linewidth, trim= 60 50 60 60 ]{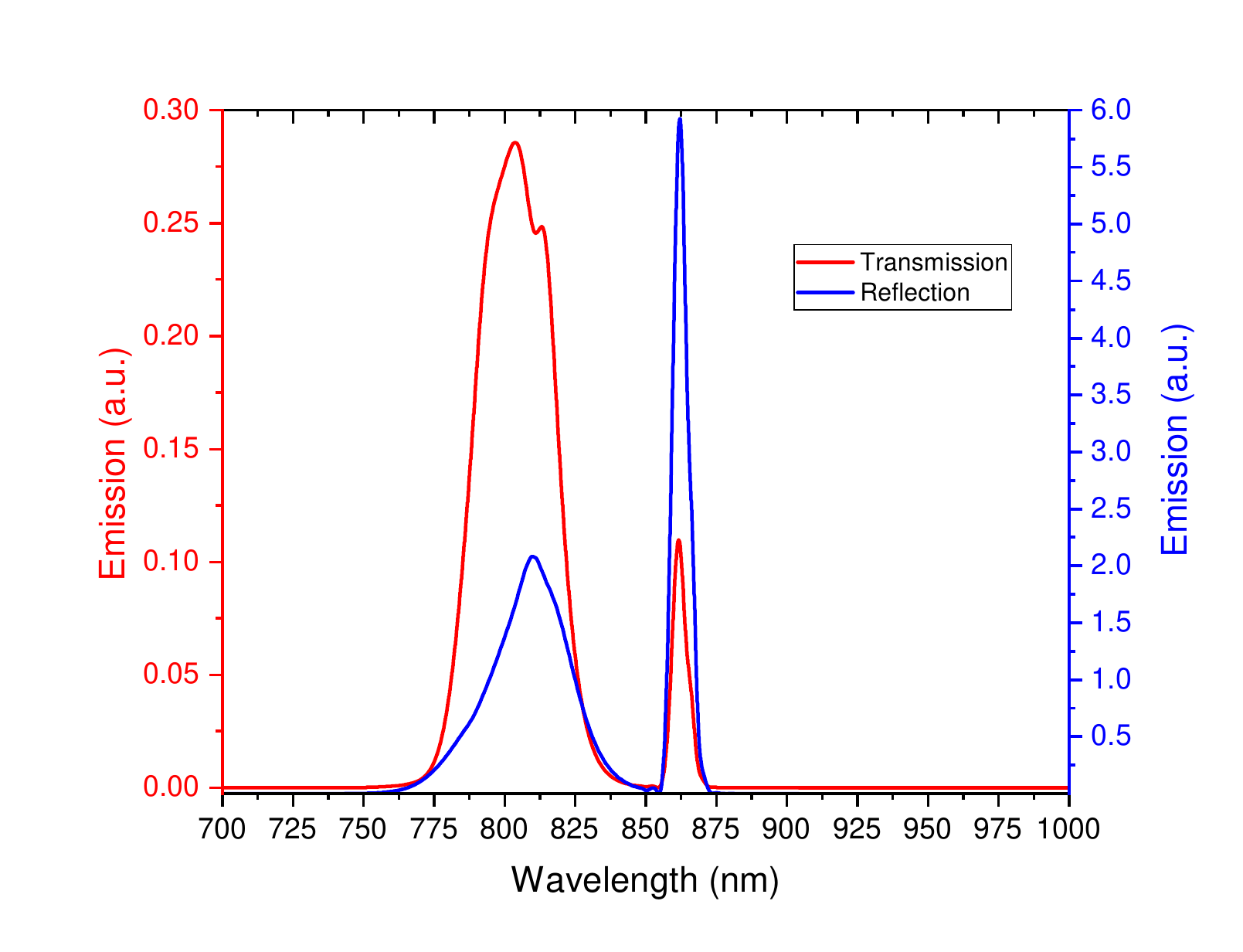}
    \captionsetup{justification=centering}
    \caption{Emission spectra at the reflection\\ and transmission side.}
    \label{T_R_noDBR}
    \end{subfigure}
    \caption{Plasmonic nanolaser cavity with no feedback mechanism and its corresponding emission spectra. The dimensions are identical to those in Figure \ref{Overall_model}.\vspace{-0.2cm}}
    \label{T_R_Pic_noDBR}
\end{figure}

   In Figure~\ref{EOT_based_cavity}, we consider an octagonal periodic nanohole array etched into a 100 nm-thick silver (Ag) slab, coated with alumina (Al\textsubscript{2}O\textsubscript{3}). A separation layer of thickness $d_\text{SL} = 20$ nm is placed between the metal layer and the IR-140-doped PU gain medium, and the structure is completed with a terminating layer of thickness $d_\text{TL} = 186$ nm. These dimensions are selected based on the parametric sweep results presented in the following sections, where we demonstrate the OTS phenomena.

   A 40~ns pump pulse of 800 nm wavelength (offset: 80~ns, amplitude: $8 \times 10^7$~V/m) excites the gain layer. As shown in Figure~\ref{T_R_noDBR}, a weak lasing peak appears at 861.658~nm on the transmission side (intensity: 0.1096~a.u.), with stronger forward scattering at 805~nm (0.2856~a.u.). Conversely, the reflection side exhibits a dominant lasing peak at 862.082~nm (5.9265~a.u.) and a forward scattering peak near 810~nm (~2~a.u.). These results indicate that most lasing energy is reflected, emphasizing the need for improved feedback to enhance transmission efficiency.

\vspace{-0.3cm}

\subsubsection{3.1.3.   Inserting the top 1DPC — OTS excitation}
\vspace{-0.2cm}

To improve the emission spectra and address the issues identified in the previous section, we propose a model that leverages Optical Tamm States (OTS) and Extraordinary Optical Transmission (EOT) phenomena. In this section, we used a plane wave source with a wavelength range of 750-950~nm to determine the optical Tamm resonance wavelength. 

     \begin{figure}[htbp]
        \centering
        \begin{subfigure}[b]{0.48\linewidth}
            \centering
            \includegraphics[width=.75\linewidth, trim= 90 50 90 80]{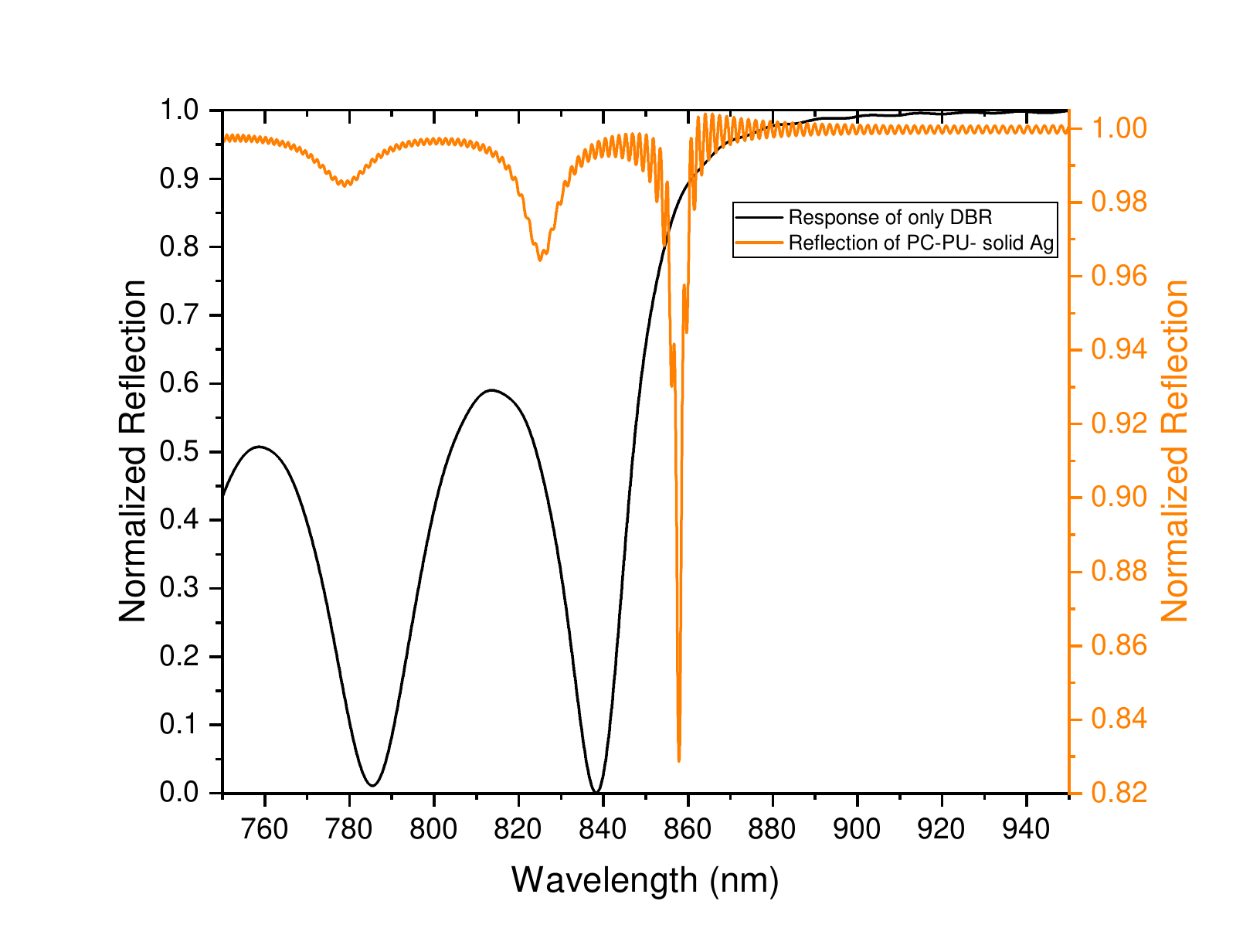}
            \captionsetup{justification=centering}
            \caption{Normalized reflection spectra for the 1DPC alone\\ and with an added PU layer capped by a Ag layer.}
            \label{OTS_view_1}
        \end{subfigure}
        \hfill
        \begin{subfigure}[b]{0.48\linewidth}
            \centering
            \includegraphics[width=.75\linewidth, trim= 90 50 90 80]{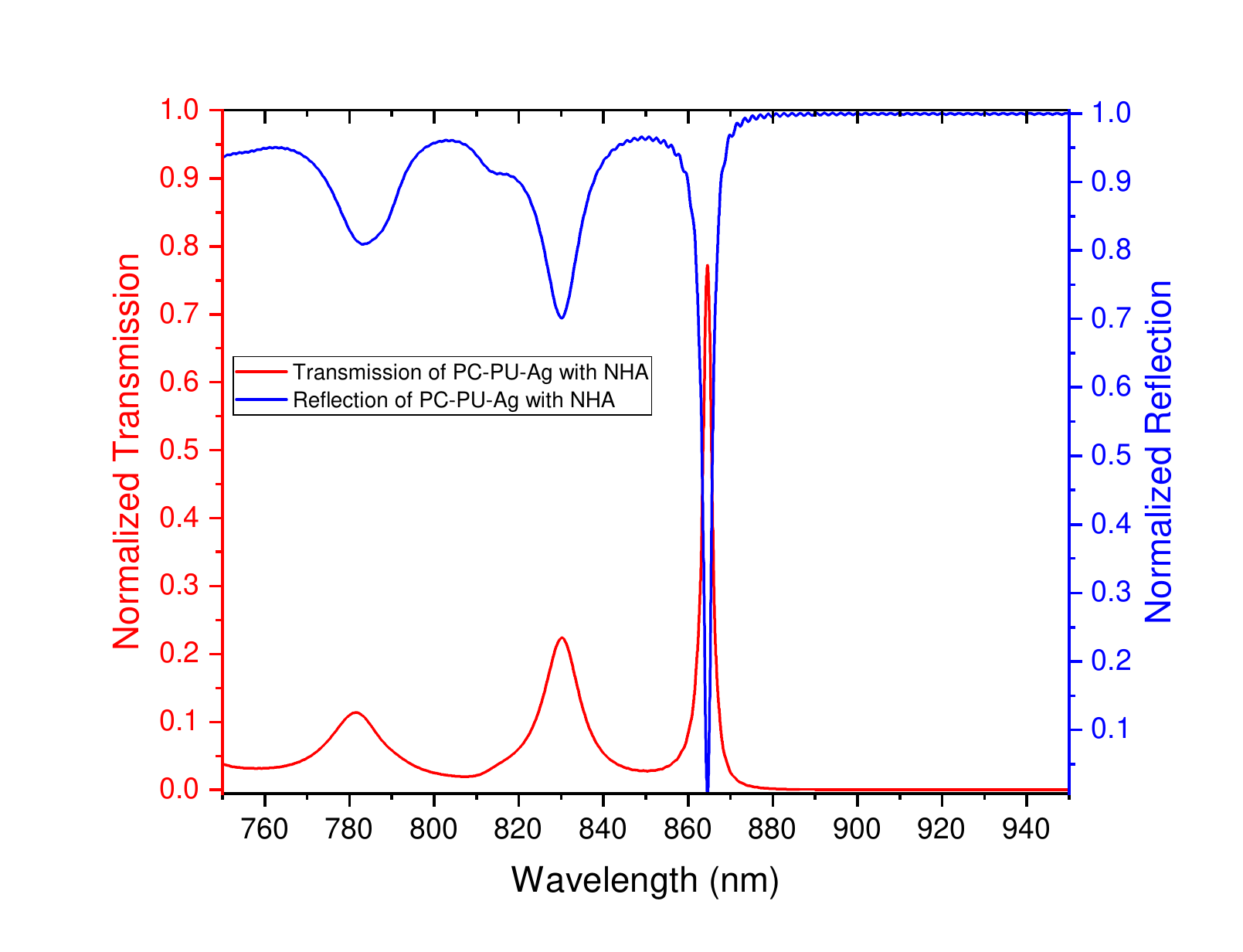}
            \captionsetup{justification=centering}
            \caption{Normalized reflection and transmission spectra\\ for the proposed PNL with perforated Ag layer.}
            \label{OTS_view_2}
        \end{subfigure}
        \caption{Normalized emission spectra of the proposed plasmonic nanolaser (PNL) observed from (a) transmission and (b) reflection sides.\vspace{-0.2cm}}
        \label{OTS_view}
    \end{figure}

When a DBR is terminated with a metal layer, an optical Tamm state (OTS) forms at certain wavelengths within the DBR stopband, trapping the electromagnetic field at the DBR-metal interface. This occurs because the phase shifts from the DBR and metal reflections match, producing constructive interference and a standing-wave mode confined at the boundary. Unlike a bare DBR with nearly 100\% reflectivity, the metal termination allows light to couple into the OTS, causing a sharp reflection dip that signifies energy confinement. The field decays inside the DBR and metal but peaks at the interface, forming a highly localized mode analogous to surface plasmons without requiring momentum matching. \cite{kavokin_optical_2005, symonds_confined_2012, treshin_optical_2013}

 Figure~\ref{OTS_view_1} shows the normalized reflection spectra for the bare top DBR only and for the top DBR terminated by a PU layer of 310 nm (host only, no dye) layer followed by a solid Ag layer. A reflection dip (in the orange curve) is observed at approximately 857.85 nm for the PC-PU-solid Ag structure, indicating the presence of an OTS resonance. This OTS resonance occurs within the stop-band region of the base DBR (PC) structure and is highly sensitive to the thicknesses of the top DBR layers, as well as the PU (host) and terminating layers (TL).  

Moreover, in Figure \ref{OTS_view_2}, the normalized transmission spectra as well as the normalized reflection of the proposed nanolaser structure, which consists of the top DBR, a PU layer of 310 nm (host only, no dye), and a perforated silver (Ag) layer, show a pronounced enhancement in extraordinary optical transmission (EOT), resulting from the OTS resonance at 864.581 nm, with normalized transmission of 0.772 and normalized reflection of 0.001851, which is really outstanding, compared to previously reported transmission of around 0.62 \cite{ahmed_efficient_2018} and a direct evidence of the coupling of EOT to OTS mode at 864.581 nm. To achieve resonant transmission close to the peak photoluminescence of the gain medium (\(\sim 870\,\text{nm}\)), the thickness of the top dielectric layer (\(\text{SiO}_2\)) is set to \(d_{\text{T1}} = 170\,\text{nm}\), the bottom layer (\(\text{TiO}_2\)) to \(d_{\text{B1}} = 111\,\text{nm}\), and the terminating layer to \(d_{\text{TL}} = 186\,\text{nm}\) after parametric sweeping and optimizations.

The coupling between optical Tamm states (OTS) and extraordinary optical transmission (EOT) occurs via two main effects. The OTS forms a surface-localized standing wave at the DBR-metal interface, enhancing the near-field and feeding the perforated holes, which excites localized and SPP-like modes contributing to transmission. Efficient coupling also requires phase- and frequency-alignment between the OTS and the natural EOT resonances of the hole array, enabling coherent far-field radiation. These effects increase transmitted intensity, sharpen spectral features, and improve outcoupling efficiency. \cite{treshin_optical_2013, kavokin_optical_2005, ebbesen_extraordinary_1998, genet_light_2007}

\vspace{-0.3cm}

\subsubsection{3.1.4.   Incorporating the gain medium with only top DBR}
\vspace{-0.2cm}

\begin{figure}[htbp]
    \centering
    % Left subfigure
    \begin{subfigure}[b]{0.48\linewidth}
        \centering
        \includegraphics[width=.8\linewidth, trim=60 10 160 30]{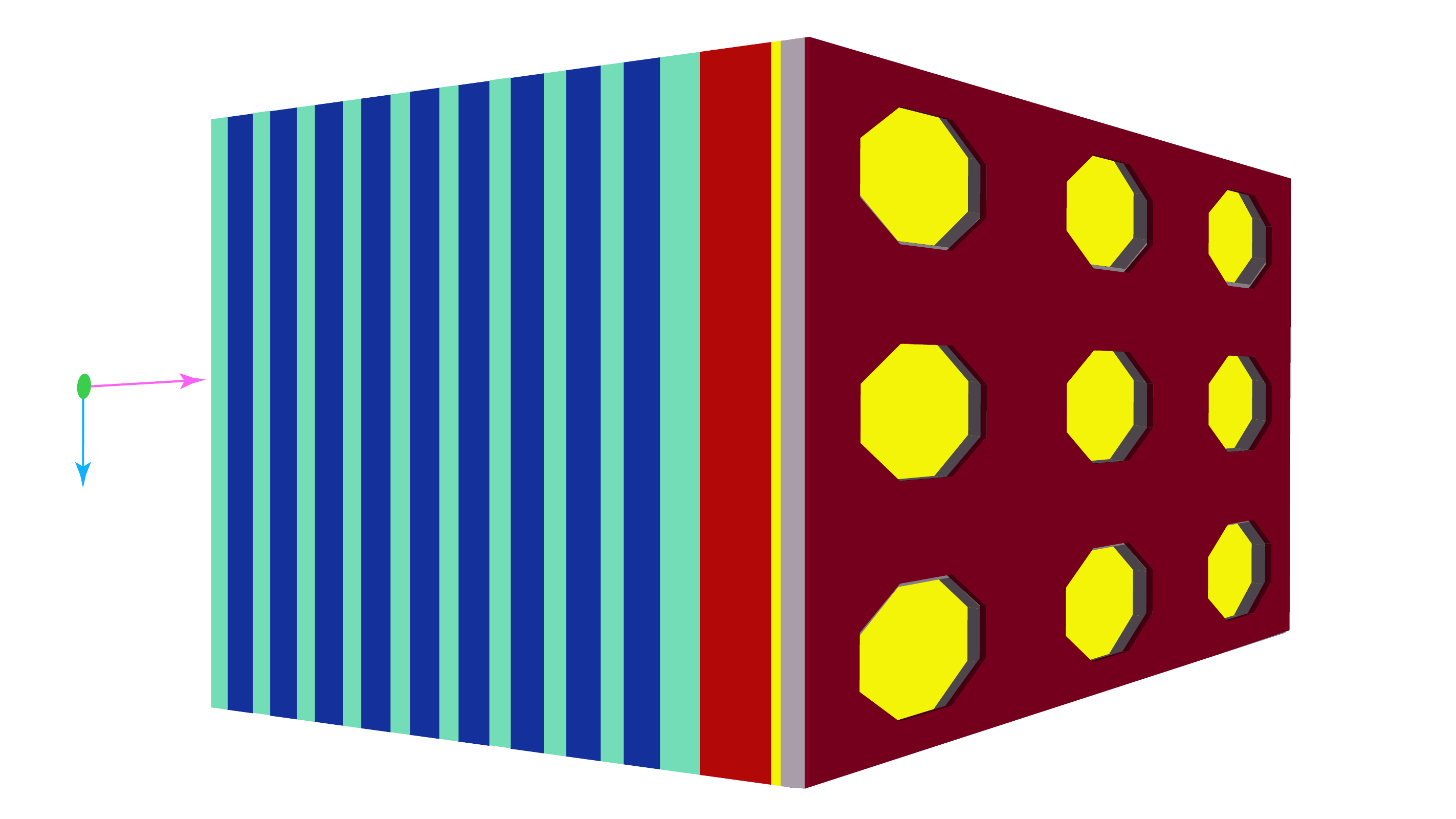}
        \caption{3D illustration of the PNL with the 1DPC at the upper side consisting of alternating layers of SiO\textsubscript{2} and TiO\textsubscript{2}.}
        \label{perspective_prev}
    \end{subfigure}
    \hfill
    % Right subfigure
    \begin{subfigure}[b]{0.48\linewidth}
        \centering
        \includegraphics[width=.8\linewidth, trim=40 0 50 140]{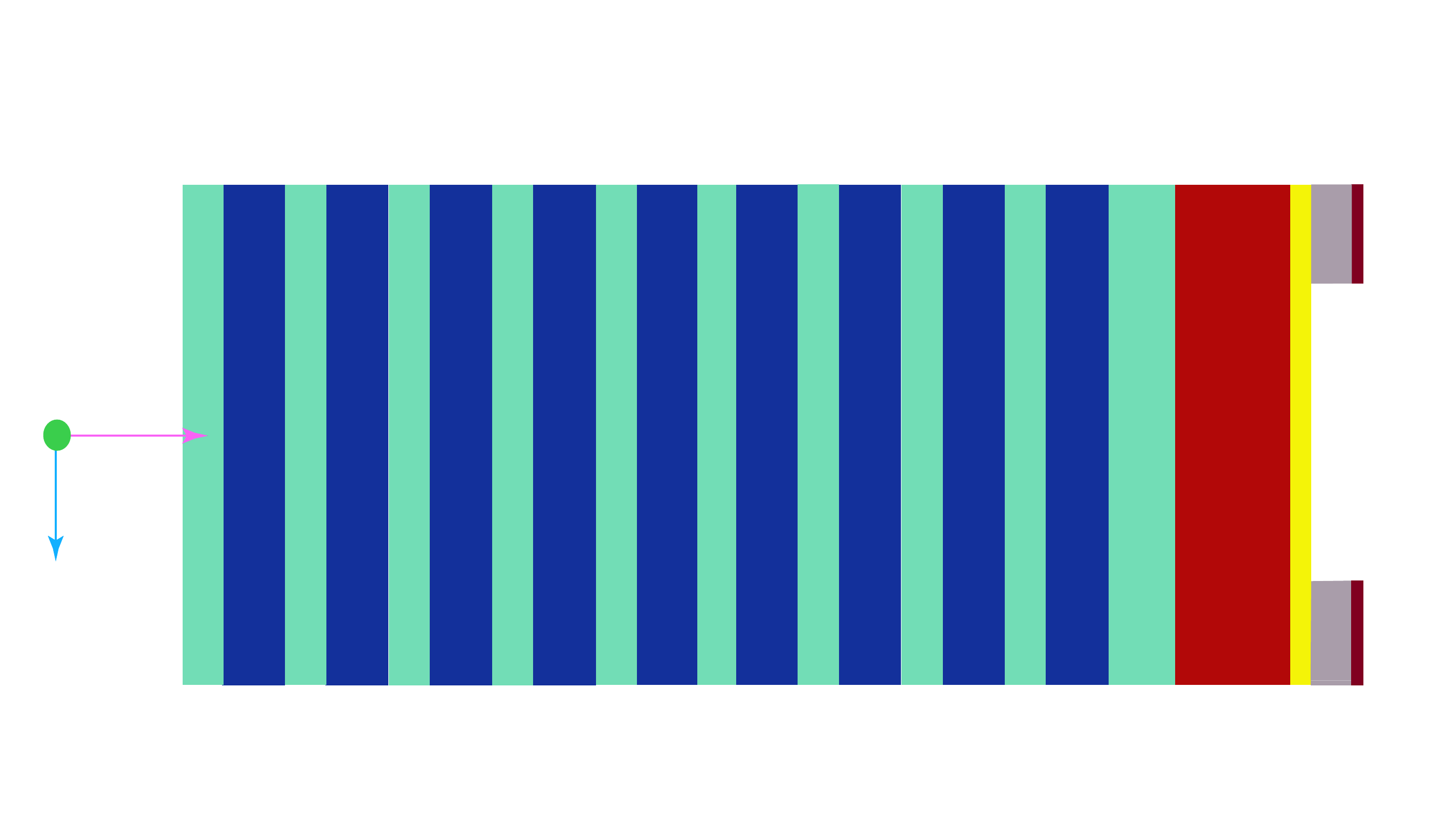}
        \caption{Cross-sectional view of a unit cell of \\the model.}
        \label{cross_section_prev}
    \end{subfigure}

    \caption{Schematic of the laser model with only top DBR.\vspace{-0.2cm}}
    \label{prev_model}
\end{figure}

The host PU layer is replaced with a gain medium by doping it with IR-140 dye as depicted in Figure \ref{prev_model}, effectively introducing optical gain into the structure. To study the lasing behavior, a pump pulse (of 800 nm) with an amplitude of \(8 \times 10^7\,\text{V/m}\), a duration of 40 fs, and an offset of 80 fs is applied. The device dimensions are the same as discussed in \ref{Design_params}.

\vspace{-0.3cm}

\subsubsubsection{A. Emission profile of the laser}

\vspace{-0.2cm}

\begin{figure}[htbp]
    \centering

    % Subfigure 1: Transmission-side spectra
    \begin{subfigure}[b]{0.3\textwidth}
        \centering
        \includegraphics[width=\textwidth, trim=110 60 60 120]{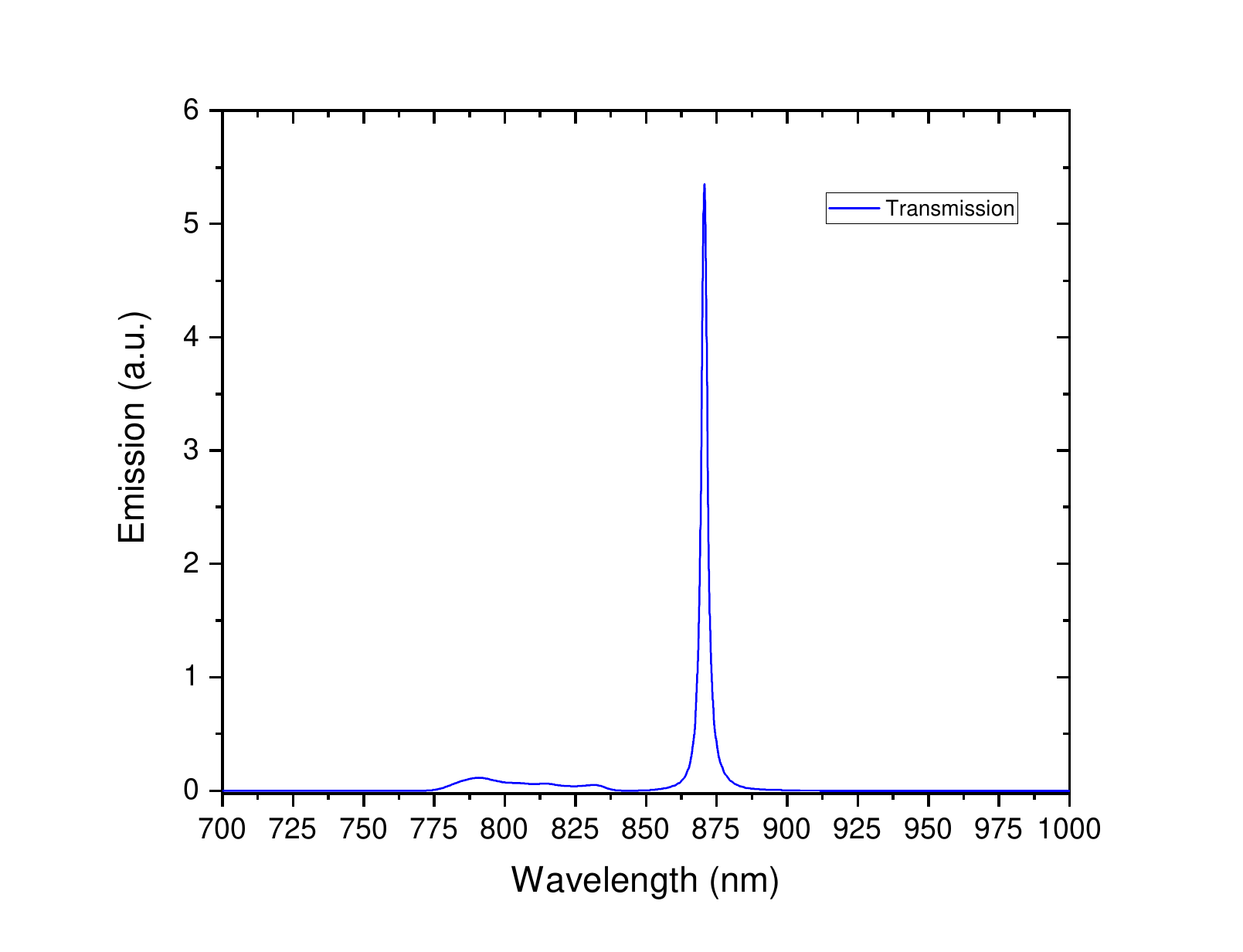}
        \captionsetup{justification=centering}
        \caption{Emission spectra in the \\transmission side.}
        \label{T_Gainadd_noDBR2}
    \end{subfigure}%
    \hspace{0.02\textwidth}%
    % Subfigure 2: Reflection-side spectra
    \begin{subfigure}[b]{0.3\textwidth}
        \centering
        \includegraphics[width=\textwidth, trim=110 60 60 100]{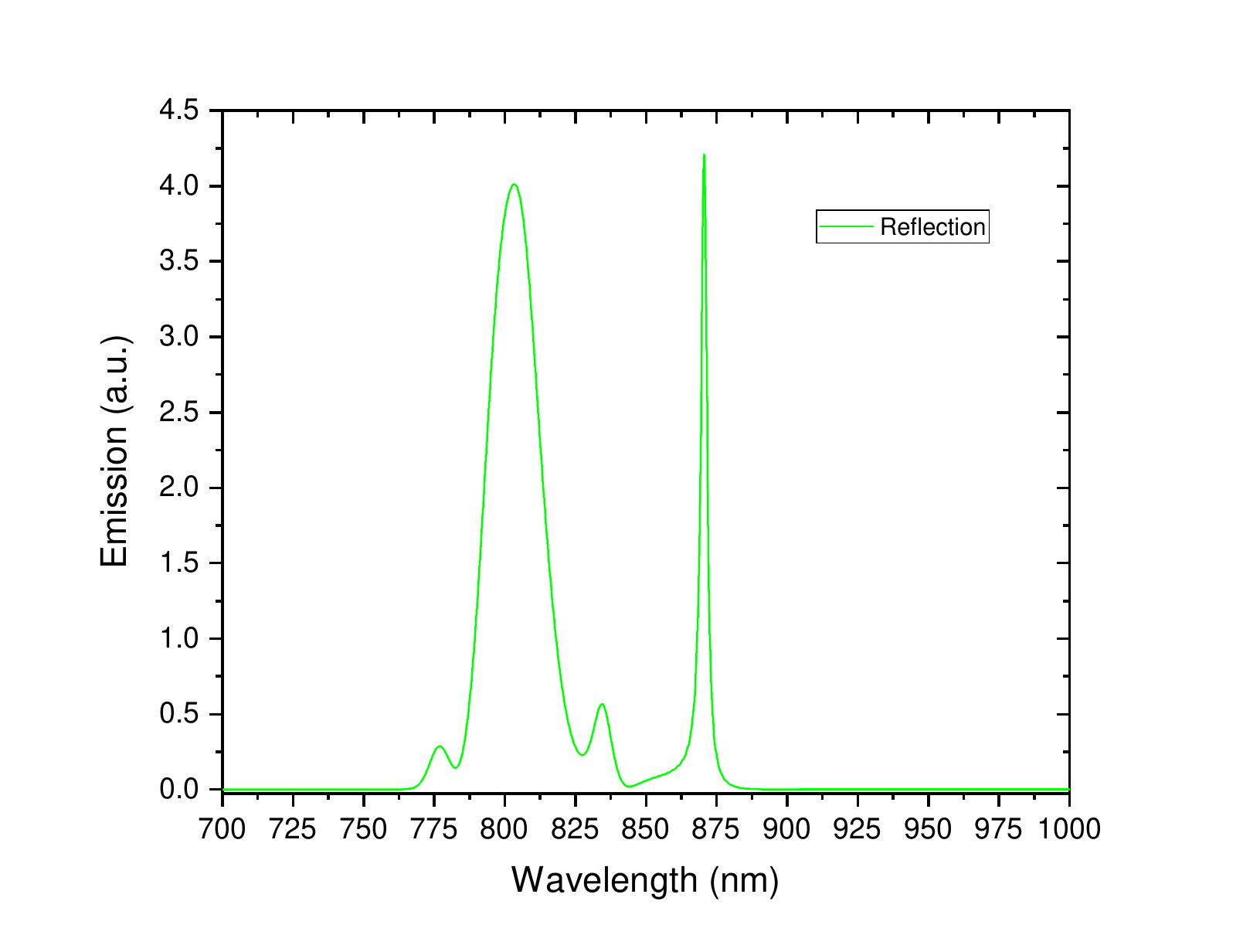}
        \captionsetup{justification=centering}
        \caption{Emission spectra in the \\reflection side.}
        \label{R_Gainadd_noDBR2}
    \end{subfigure}%
    \hspace{0.02\textwidth}%
    % Subfigure 3: Pump amplitude variation
    \begin{subfigure}[b]{0.33\textwidth}
        \centering
        \includegraphics[width=\textwidth, trim=40 40 25 55,clip]{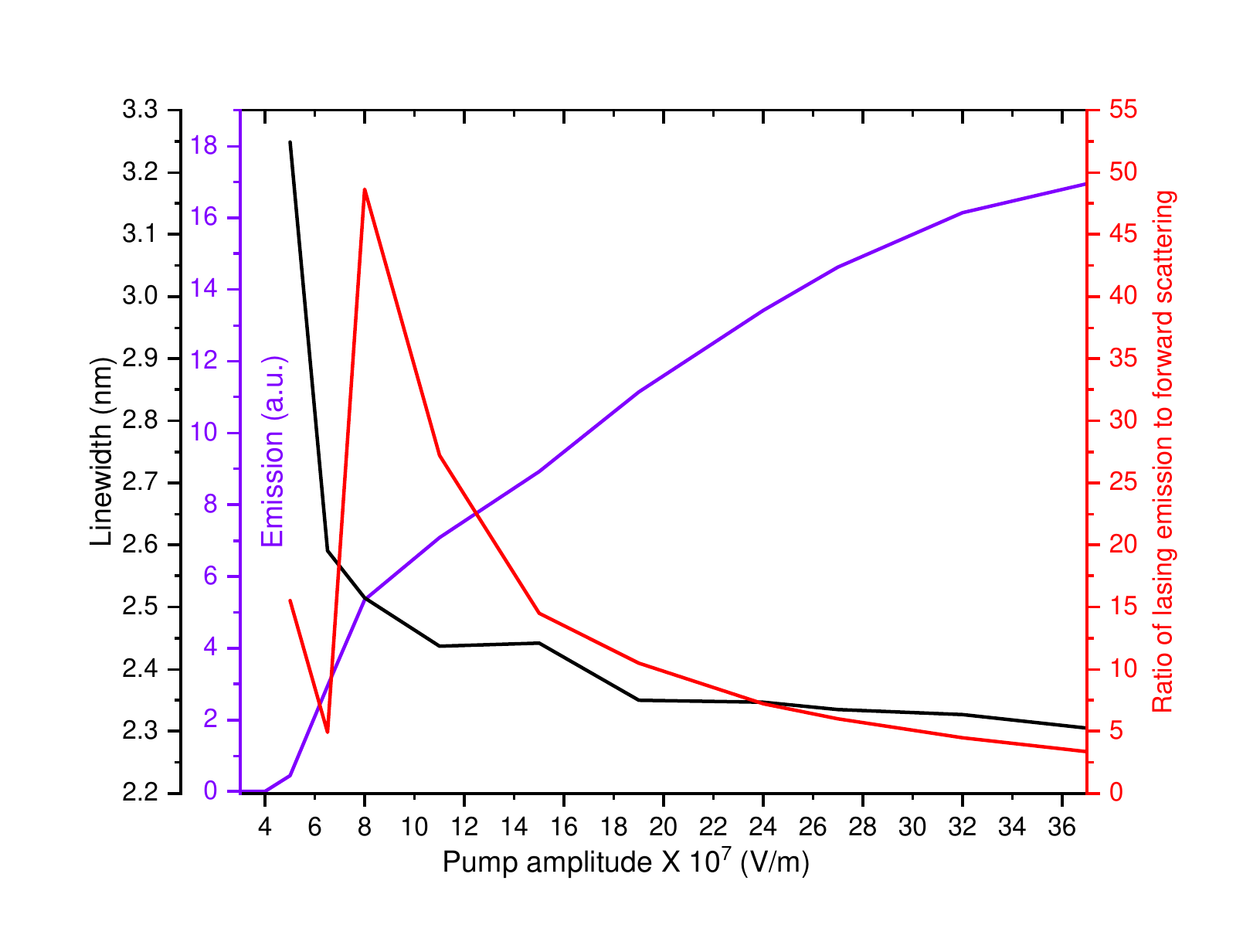}\captionsetup{justification=centering}
        \caption{Variation of various emission\\ characteristics with pump amplitude.}
        \label{amp_varied}
    \end{subfigure}

    \caption{Analysis of emission characteristics of the single-DBR PNL model.\vspace{-0.3cm}}
    \label{PNL_emission_3sub}
\end{figure}

Lasing emission (a.u.) appears on both transmission and reflection sides, as shown in Figures~\ref{T_Gainadd_noDBR2} and \ref{R_Gainadd_noDBR2}. On the transmission side, lasing occurs at 870.055~nm with an intensity of 5.351~a.u., accompanied by pump scattering near 800~nm (0.1116~a.u.). The reflection side shows notable emission at both the lasing (4.206~a.u.) and pump (4.010~a.u.) wavelengths, indicating partial leakage due to nonideal 1D~PC reflectivity.

Before introducing the dye, the OTS resonance determined solely by the DBR-metal structure occurs at 864.581~nm (Figure~\ref{OTS_view_2}). Incorporating the dye increases the effective refractive index of the host medium, slightly extending the optical path and redshifting the resonance to 870.055~nm. As the dye’s gain spectrum also peaks near 870~nm, the lasing emission naturally aligns with this wavelength.

\vspace{-0.3cm}
\subsubsubsection{B. Pump amplitude selection}

\vspace{-0.2cm}

We now proceed to examine from Figure \ref{amp_varied} how variations in the input pump pulse amplitude influence the emission characteristics on the transmission side, including the intensity at the lasing wavelength, the emission linewidth and the ratio of lasing emission to the forward scattering of the pump wavelength.

Figure \ref{amp_varied} shows that for pump amplitudes below \(5 \times 10^{7}\,\text{V/m}\) (corresponding to \(0.01\,\text{mJ/cm}^{2}\) of pump energy), no lasing occurs in the device (purple curve). As the pump amplitude increases beyond \(5 \times 10^{7}\,\text{V/m}\), the emission intensity at the lasing wavelength eventually saturates. In contrast, the linewidth (black curve) exhibits minimal improvement and appears nearly saturated for pump amplitudes above \(8 \times 10^{7}\,\text{V/m}\) (\(0.0256\,\text{mJ/cm}^{2}\) of pump energy).

In laser systems, output power initially increases with pump amplitude as higher pumping enhances population inversion and stimulated emission. However, at high pump levels, gain saturation and inversion clamping occur since the upper level depletes rapidly. Consequently, further pumping yields little additional output, with excess energy dissipated as heat, nonradiative recombination, or nonlinear losses.

Additionally, the ratio of lasing emission to forward-scattered pump light (red curve) shows a non-monotonic trend. Instead of continuously increasing with pump amplitude, the ratio reaches a maximum at \(8 \times 10^{7}\,\text{V/m}\). Beyond this point, it decreases sharply, indicating that pump scattering begins to compete with lasing emission, which is undesirable for efficient operation. Hence, a pump amplitude of \(8 \times 10^{7}\,\text{V/m}\) (\(0.0256\,\text{mJ/cm}^{2}\) of pump energy) is chosen as the optimal condition. Choosing the novel octagonal NHA results in a lower pump threshold amplitude compared to $1 \times 10^{8}\,\text{V/m}$ reported for single Tamm lasing in similar work.\cite{ahmed_efficient_2018}

\vspace{-0.3cm}

\subsubsubsection{C. Achieving population inversion}
\vspace{-0.2cm}

\begin{figure}[htbp]
        \centering
        \begin{subfigure}[b]{0.48\linewidth}
            \centering
            \includegraphics[width=.62\linewidth, trim= 90 50 120 80]{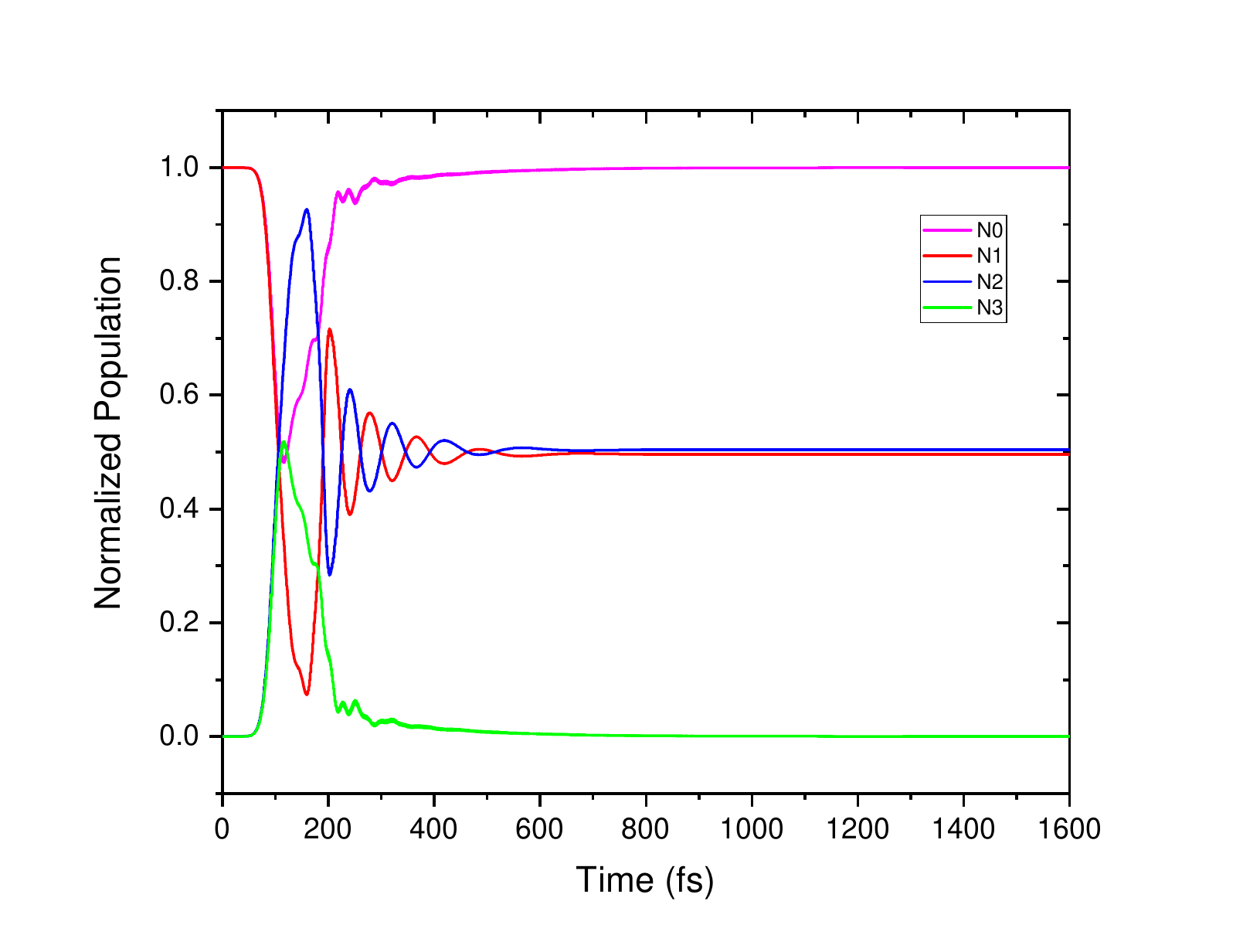}
            \captionsetup{justification=centering}
            \caption{Population density of all four levels.}
            \label{Pop_plots_no_DBR2}
        \end{subfigure}
        \hfill
        \begin{subfigure}[b]{0.48\linewidth}
            \centering
            \includegraphics[width=.62\linewidth,  trim= 90 50 120 80]{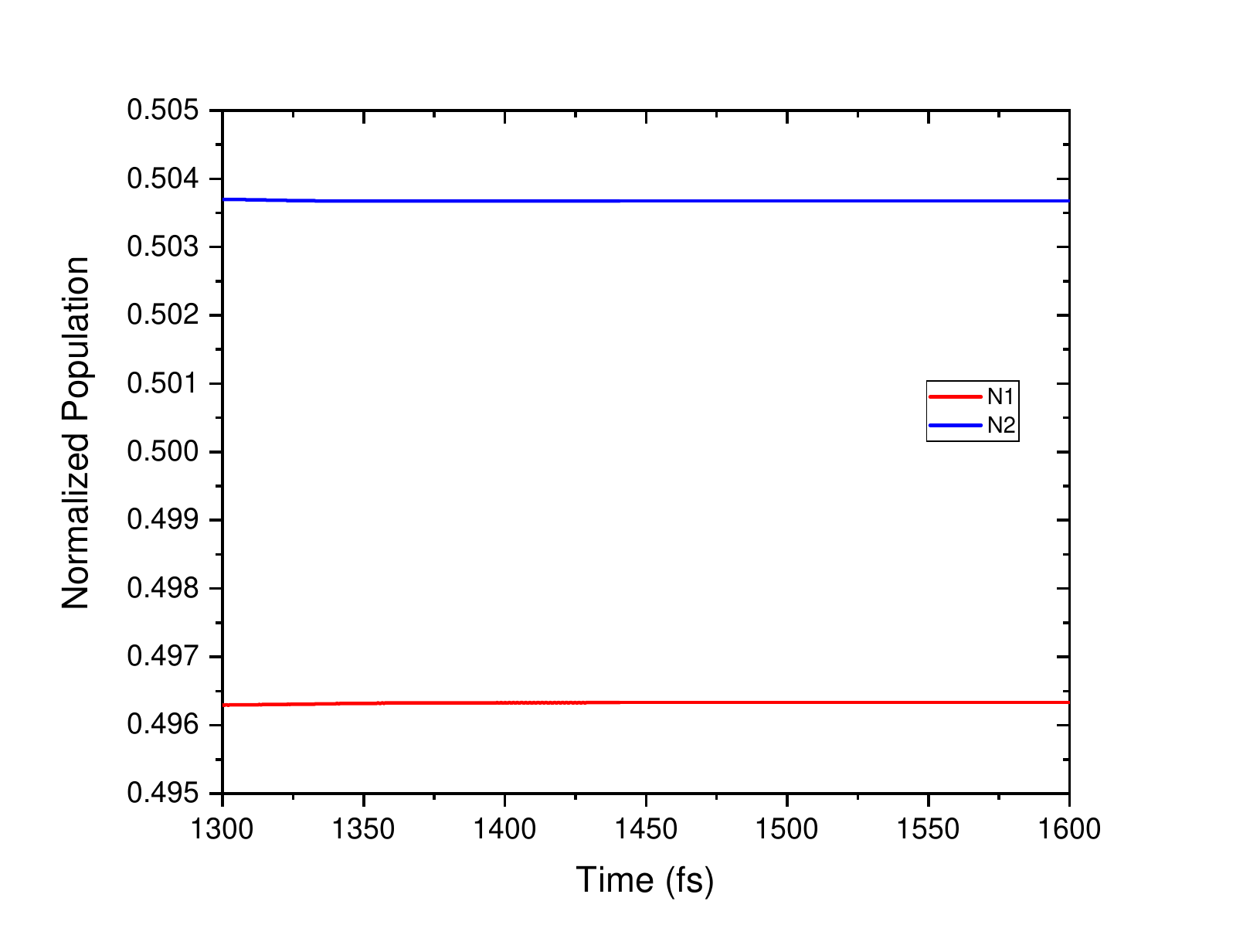}
            \captionsetup{justification=centering}
            \caption{Population density of only lasing levels.}
            \label{N1N2_only_no_DBR2}
        \end{subfigure}
        \caption{Time evolution of normalized population densities for pump amplitude of $8 \times 10^7$ V/m.\vspace{-0.62cm}}
        \label{Pop_plots_noDBR2}
    \end{figure}

Figure~\ref{Pop_plots_no_DBR2} shows the time evolution of the normalized population densities of $N_0$ (ground level), $N_1$, $N_2$, and $N_3$, while Figure~\ref{N1N2_only_no_DBR2} focuses on $N_1$ and $N_2$, allowing us to visualize the population inversion at steady state. Population inversion is a prerequisite for lasing action. We observe that the difference between the $E_2$ and $E_1$ levels stabilizes at steady state ($\Delta N / N0 $ is around $0.008034$) after some initial oscillations when the gain medium is excited by the pump pulse.

\subsubsection{3.1.5.   Issues pertaining to the current design}

\vspace{-0.2cm}

After introducing the gain medium with only the top DBR, the transmission-side emission spectra (Figure~\ref{T_Gainadd_noDBR2}) show an intensity ratio of \(5.351 / 0.1116 \approx 47.94\) between the lasing peak at 870~nm and the forward-scattered pump near 800~nm, confirming strong lasing dominance. On the reflection side (Figure~\ref{R_Gainadd_noDBR2}), the corresponding ratio of \(4.2057 / 4.01028 \approx 1.05\) indicates substantial pump scattering, making this side unsuitable for lasing. The integrated transmitted and reflected powers are 77.297 and 66.1545~a.u., respectively. Both emission peak at lasing intensity and integrated power show that a significant fraction of lasing energy escapes through reflection side, a limitation of the current design.

\vspace{-0.3cm}

\subsubsection{3.1.6.   Proposing an improved design : A Dual-DBR Model}

\vspace{-0.2cm}

To enhance the optical feedback of the nanocavity and reduce lasing energy escaping through the reflection side, we introduce a second (bottom) DBR composed of alternating layers of MgF\textsubscript{2} (\(d_{\text{T2}} = 60\,\text{nm}\)) and TiO\textsubscript{2} (\(d_{\text{B2}} = 140\,\text{nm}\)), designed to have a stopband around 870 nm. The redesigned structure has already been shown in Figure~\ref{Overall_model}. The pump characteristics remain unchanged.

\vspace{-0.3cm}

\subsubsubsection{A. Emission spectra of the Dual-DBR Model}

\vspace{-0.2cm}
We now analyze the emission characteristics of the improved design and compare them with the previous structure. First, we consider the transmission and reflection sides.

     \begin{figure}[htbp]
    \centering
    \begin{subfigure}[b]{0.48\linewidth}
        \centering
        \includegraphics[width=.7\linewidth, trim= 90 30 80 80 ]{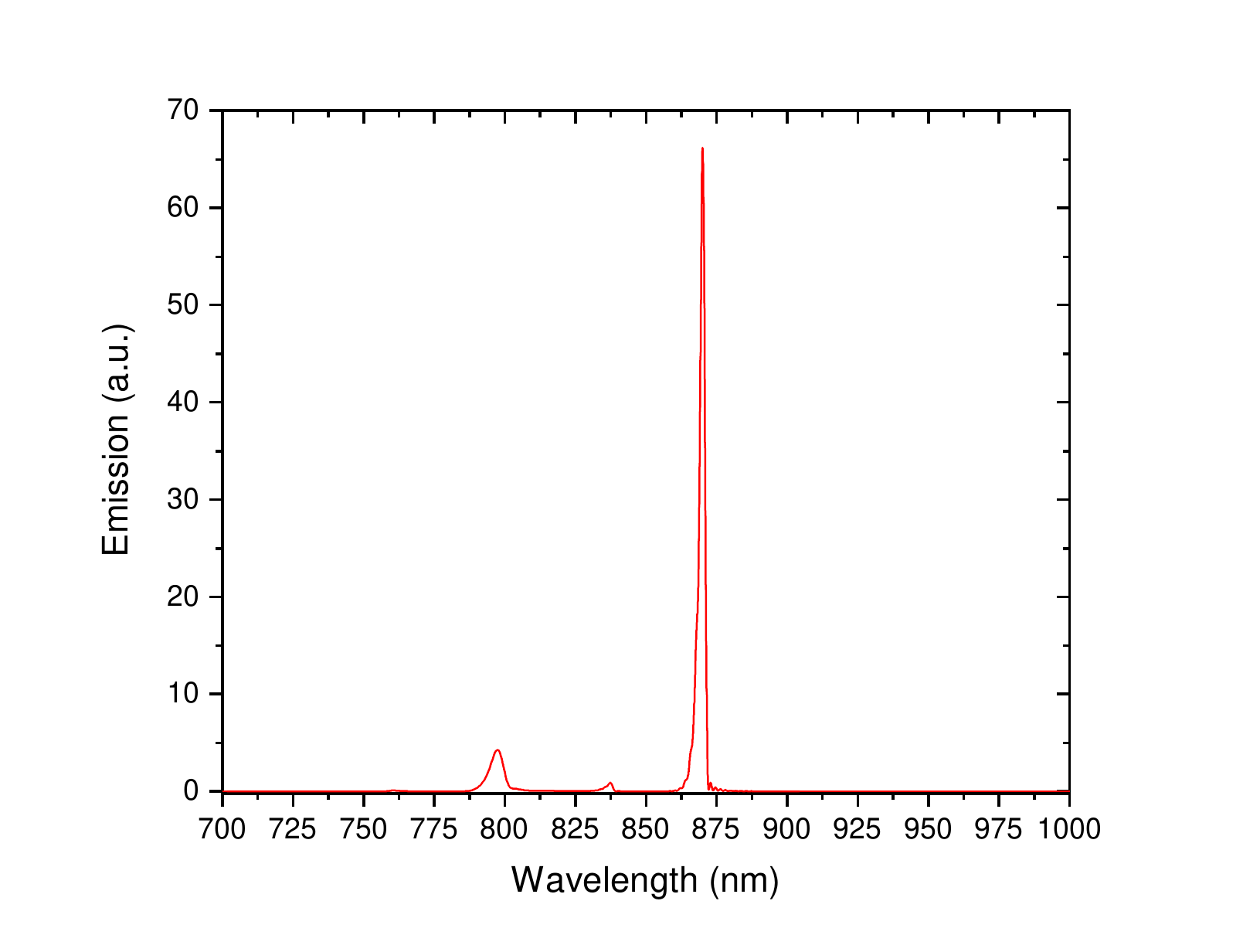}
        \caption{Transmission-side emission spectra}
        \label{DBR2added_T}
    \end{subfigure}
    \hfill
    \begin{subfigure}[b]{0.48\linewidth}
        \centering
        \includegraphics[width=.7\linewidth, trim= 110 30 70 80]{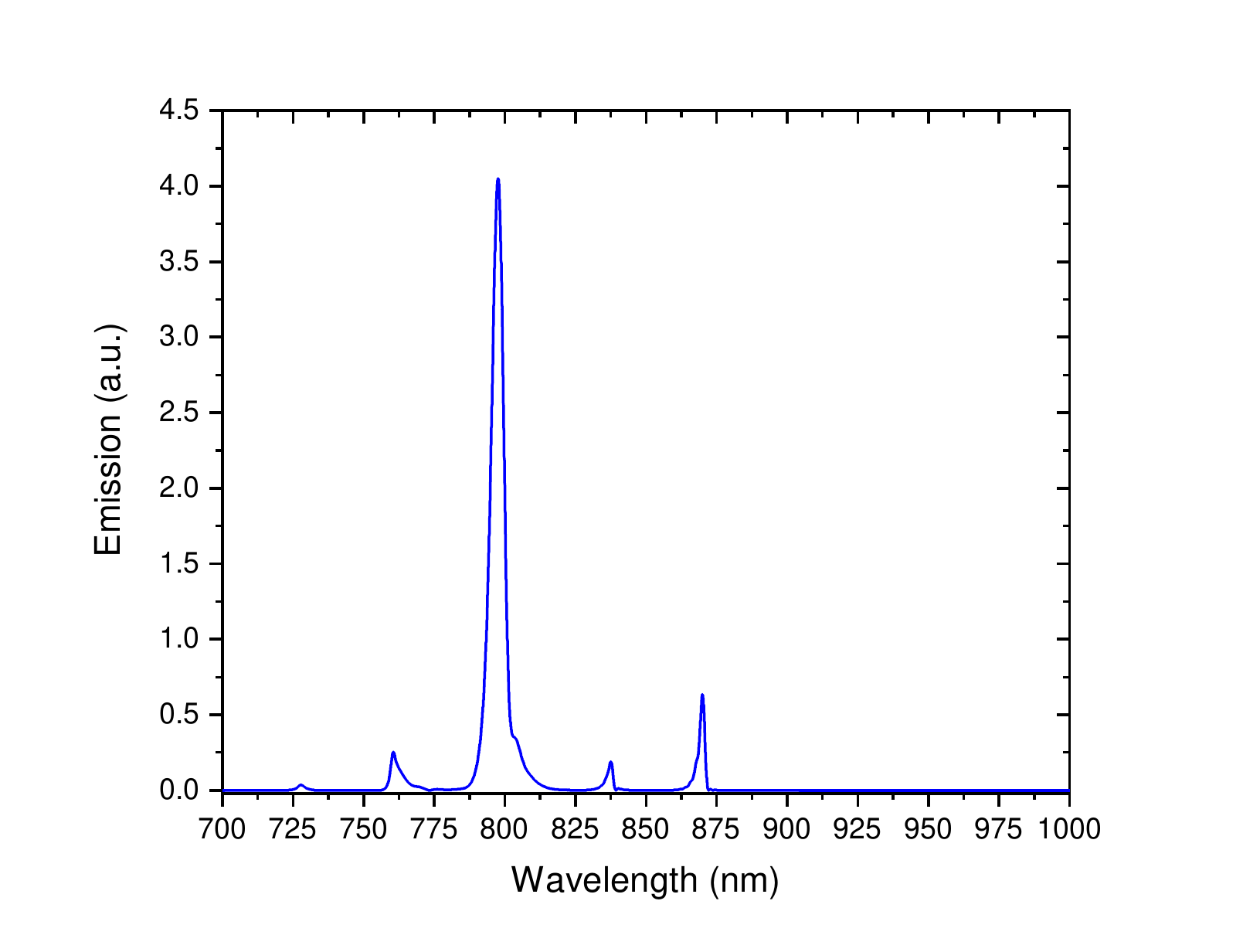}
        \caption{Reflection-side emission spectra}
        \label{DBR2added_R}
    \end{subfigure}
    \caption{Emission spectra of the final PNL model proposed in Figure \ref{Overall_model}, observed from (a) transmission and (b) reflection side.\vspace{-0.3cm}}
    \label{PNL_emission_spectra2}
\end{figure}

 In Figure \ref{DBR2added_T} on the transmission side, the peak emission occurs at 870.055 nm with an intensity of 66.203 a.u., while the emission on the reflection side (in Figure \ref{DBR2added_R}) at the same wavelength is reduced to only 0.623 a.u., representing an approximately 106-fold higher lasing intensity. The computed integrated emission power in the transmission side is around 664.708 a.u., compared to 6.889 a.u. on the reflection side, corresponding to nearly a 96.5-fold increase in transmitted power relative to the reflected power.

Relative to the previous design without the bottom DBR, the transmission-side peak intensity shows a 12.37-fold improvement (\(66.203/5.351\)), and the integrated emission power rises by a factor of 8.6 (\(664.708/77.297\)) for the same input pump energy. These enhancements are achieved because the bottom DBR now efficiently reflects the escaping lasing wavelength (870 nm) back, redirecting it toward the transmission side.

\vspace{-0.3cm}

\subsubsubsection{B. Population inversion profile}

\vspace{-0.2cm}

\begin{figure}[htbp]
        \centering
        \begin{subfigure}[b]{0.48\linewidth}
            \centering
            \includegraphics[width=.62\linewidth, trim= 90 50 140 90]{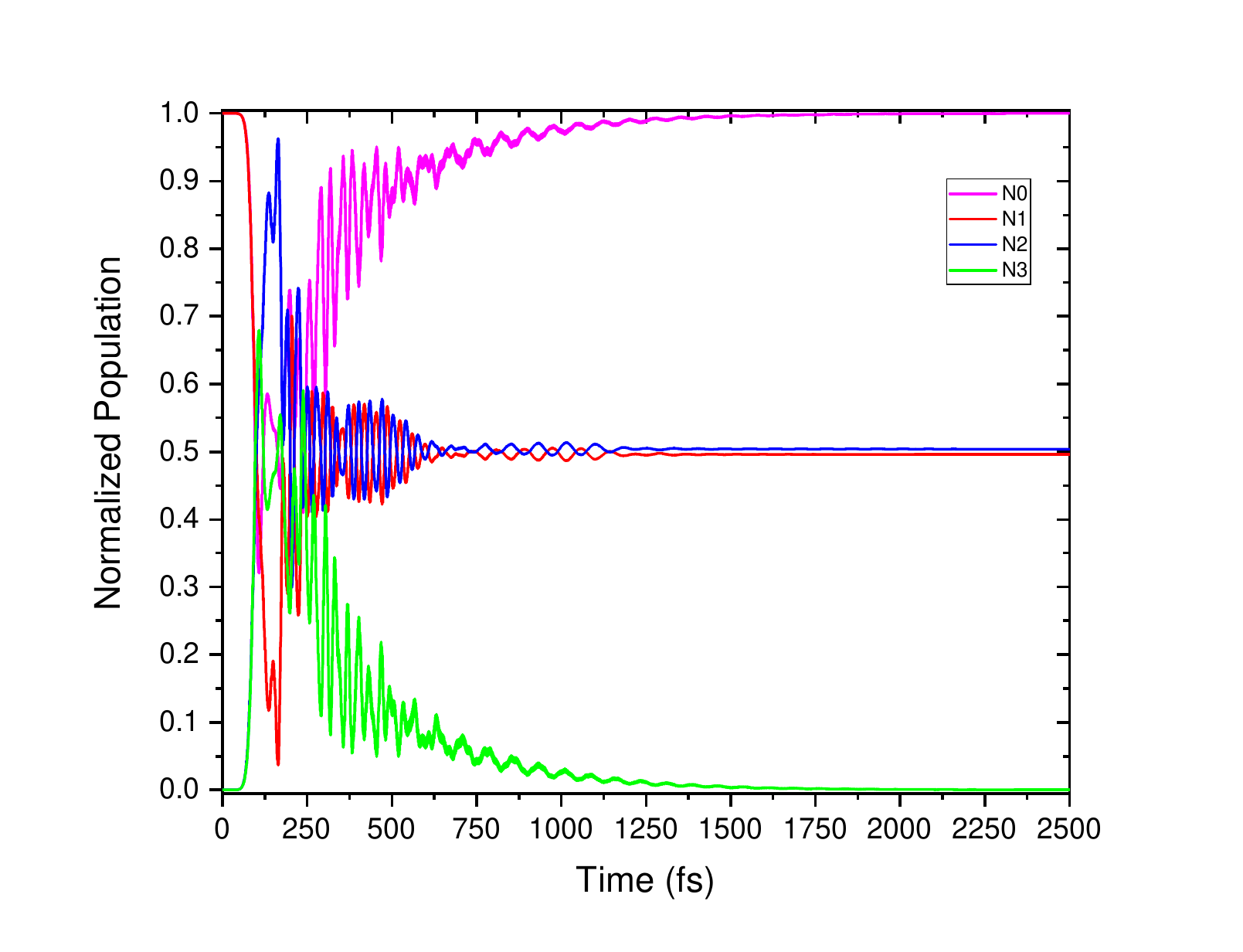}
            \caption{Population density of all four levels.}
            \label{DBR2_pop}
        \end{subfigure}
        \hfill
        \begin{subfigure}[b]{0.48\linewidth}
            \centering
            \includegraphics[width=.62\linewidth, trim= 90 50 140 90]{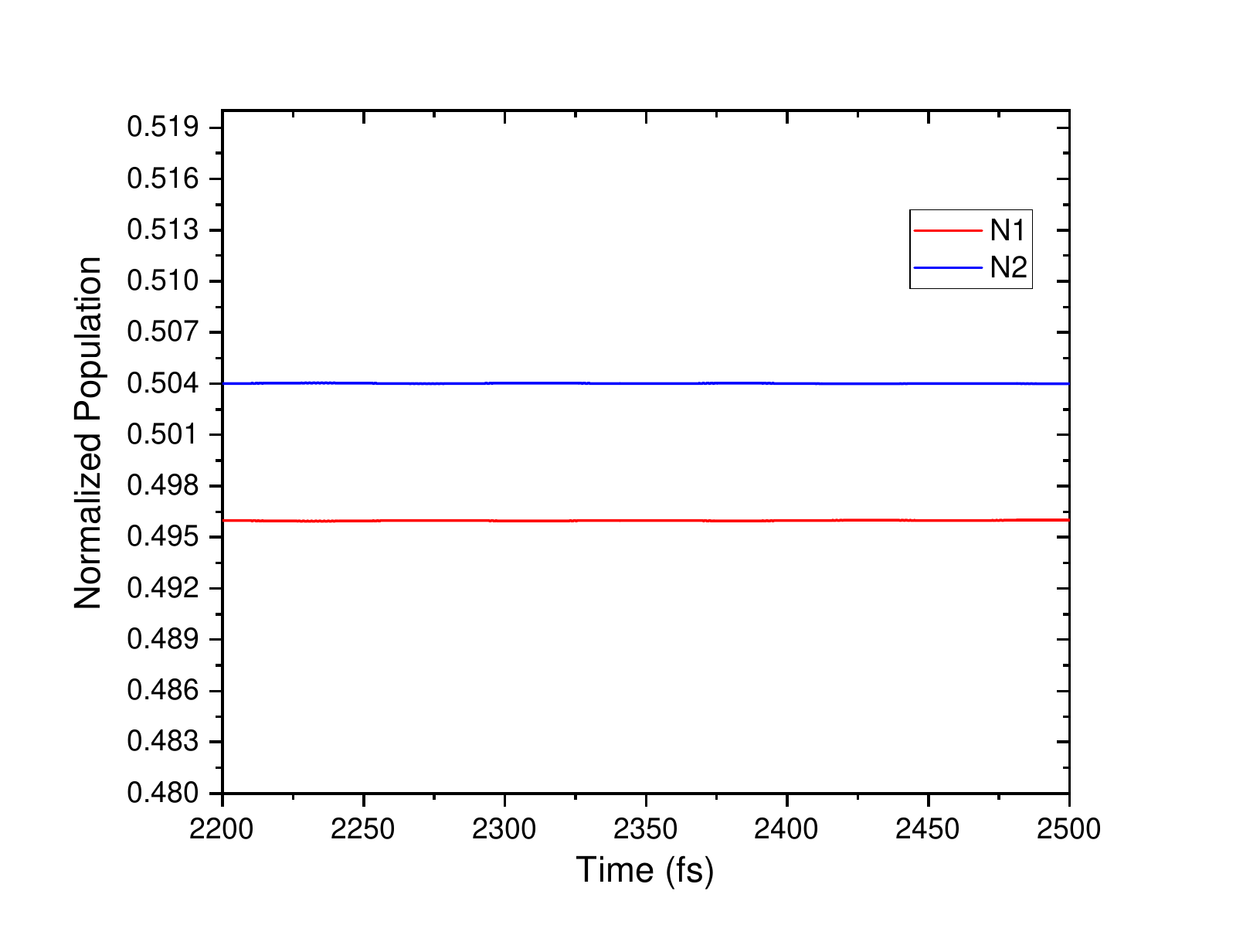}
            \caption{Population density of only lasing levels.}
            \label{DBR2_N1N2}
        \end{subfigure}
        \caption{Time evolution of normalized population densities for pump amplitude of $8 \times 10^7$ V/m.\vspace{-0.3cm}}
        \label{DBR2_pop_}
    \end{figure}

Analyzing Figure~\ref{DBR2_pop_}, compared to the normalized population inversion, $\Delta N / N0 = 0.008034$ approximately as seen in the model with only the upper 1DPC in Figure \ref{N1N2_only_no_DBR2}, this figure now appears to be around $0.007349$ after adding the bottom DBR. As a result, our newly designed model consumes, and hence requires less pump power to maintain the inversion between two lasing levels, thus enabling us to implement a low-power-consumption laser.

\vspace{-0.3cm}

\subsubsubsection{C. Electric field intensity distribution profile}

\vspace{-0.2cm}

\begin{figure}[htbp]
    \centering
    \includegraphics[width=0.5\linewidth, trim=70 250 70 260]{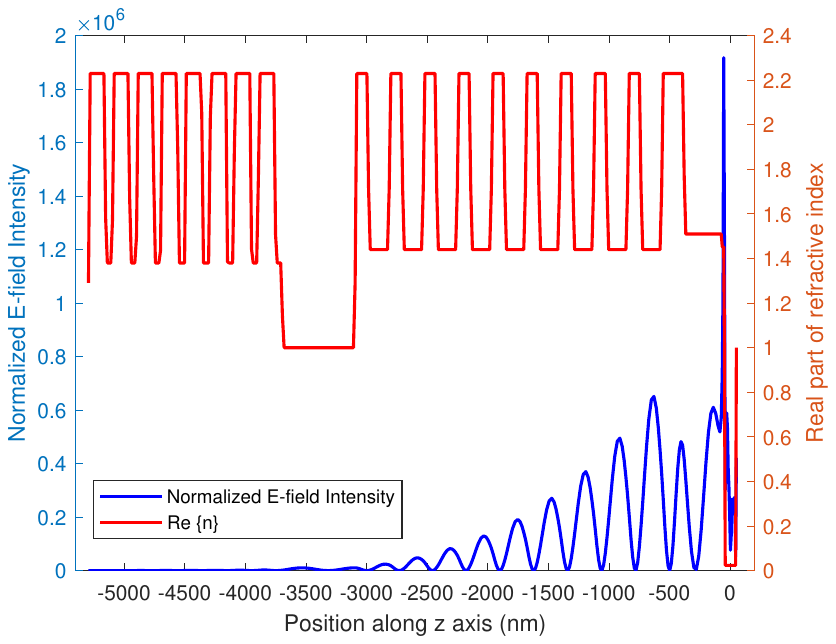}
    \caption{Refractive index profile and normalized electric field intensity distribution along the device z-axis, measured at the monitor placed on the right straight arm of the structure.\vspace{-0.4cm}}
    \label{RefIndex_Efield_at_holEdge}
\end{figure}

Figure~\ref{RefIndex_Efield_at_holEdge} presents the refractive index distribution and normalized electric field along the vertical axis at the lasing wavelength of 870.055~nm, measured from the monitor at the right arm. The electric field reaches its maximum of \(1.91745 \times 10^{6}\) (normalized) at 53.8611~nm below the perforated metal layer, which is about \(1.614423 \times 10^{6}\) times higher than the free-space value of 1.1877, showing very strong field confinement in metal-separation layer interface due to OTS. 

\begin{figure}[htbp]
    \centering

    % Top row: a, c, e
    \begin{subfigure}[b]{0.32\linewidth}
        \centering
        \includegraphics[width=\linewidth, trim=100 230 123 240,clip]{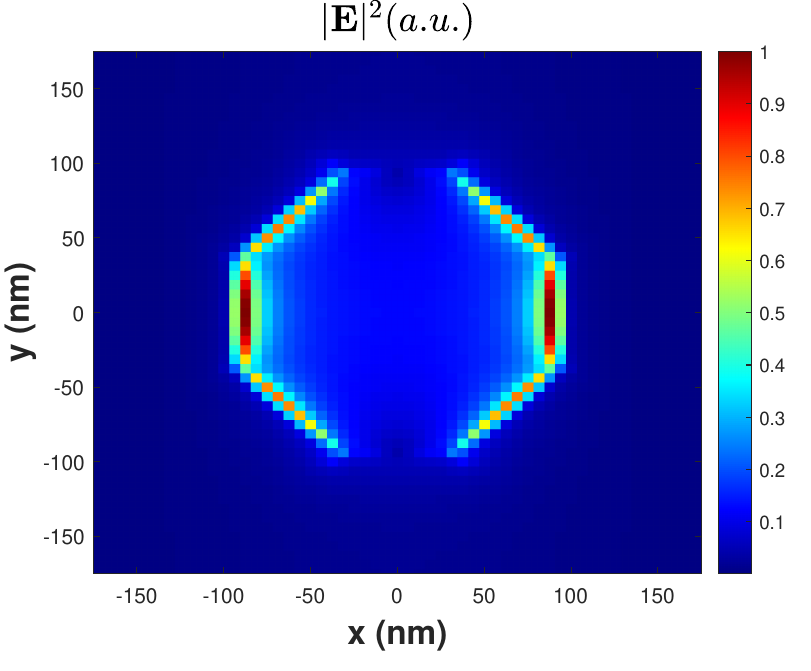}
        \captionsetup{justification=centering}
        \caption{Intensity profile for PNL \\without DBR.}
        \label{Field_xy_noDBR}
    \end{subfigure}
    \hfill
    \begin{subfigure}[b]{0.32\linewidth}
        \centering
        \includegraphics[width=\linewidth, trim=100 230 123 240,clip]{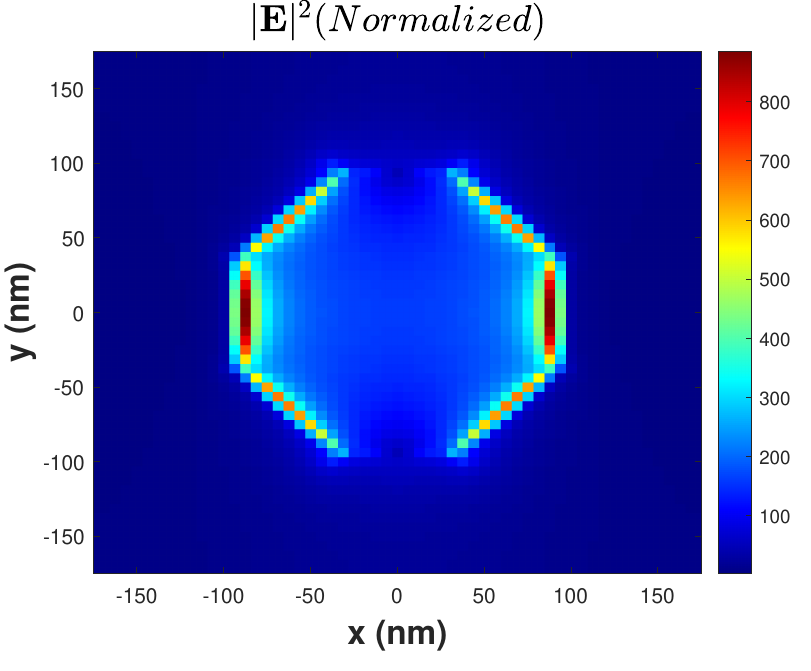}
        \captionsetup{justification=centering}
        \caption{Intensity profile for single\\-DBR PNL model.}
        \label{Field_xy_noDBR2}
    \end{subfigure}
    \hfill
    \begin{subfigure}[b]{0.32\linewidth}
        \centering
        \includegraphics[width=\linewidth, trim=100 230 123 240,clip]{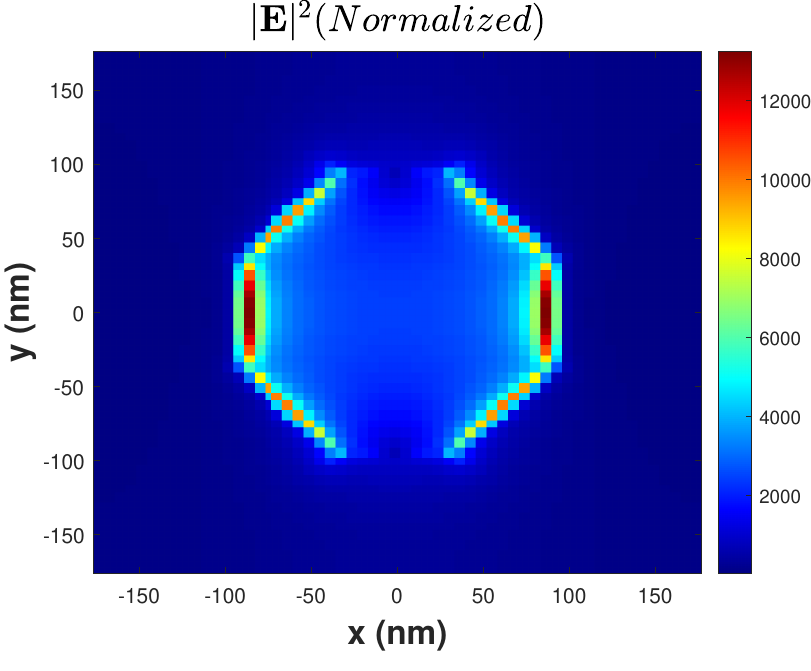}
        \captionsetup{justification=centering}
        \caption{Intensity profile for dual\\-DBR PNL model.}
        \label{Field_xy_DBR2}
    \end{subfigure}

    \caption{Electric field intensity distributions at the lasing wavelength for: (a) the bare plasmonic NHA model, (b) the initial PNL design with top DBR only, and (c) the final design with the bottom DBR incorporated. Note that (b) and (c) are normalized to the maximum intensity obtained for the bare plasmonic structure.\vspace{-0.05cm}}
    \label{Intensity_compare}
\end{figure}

We thus analyze the spatial electric field distribution to understand the lasing mechanism and the role of the 1D photonic crystal (PC) in enhancing performance, as shown in Figure~\ref{Intensity_compare}. In the bare NHA, intensity enhancement occurs at the edges of the octagonal holes due to localized surface plasmon (LSP) modes (Figure~\ref{Field_xy_noDBR}). Figures~\ref{Field_xy_noDBR2} and \ref{Field_xy_DBR2} show field intensity along the metal-separation layer interface for the top-DBR-only and dual-DBR structures, respectively. The bare structure without 1D PCs has extremely weak intensity-about 885 times weaker than the top-DBR-only case and nearly $1.3 \times 10^{4}$ times weaker than the dual-DBR design. Adding the bottom DBR increases intensity by $\sim14.7$ times over the top-DBR-only structure. Such enhancement indicates strong coupling of stimulated emission to OTS modes, improving confinement and plasmonic interaction. The $1.3 \times 10^{4}$-fold increase with dual DBRs exceeds prior reports such as Zabir \textit{et al.}~\cite{ahmed_efficient_2018} that showed only $\sim50$ times enhancement, and Figures~\ref{Field_xy_DBR2} confirms that dual DBRs produce the strongest field enhancement and confinement, validating their effectiveness in boosting nanolaser performance.

\vspace{-0.3cm}

\subsubsubsection{D. Threshold reduction}
\vspace{-0.2cm}

\begin{figure}[htbp]
    \centering
    \begin{subfigure}[b]{0.48\linewidth}
        \centering
        \includegraphics[width=.7\linewidth, trim= 35 30 160 90]{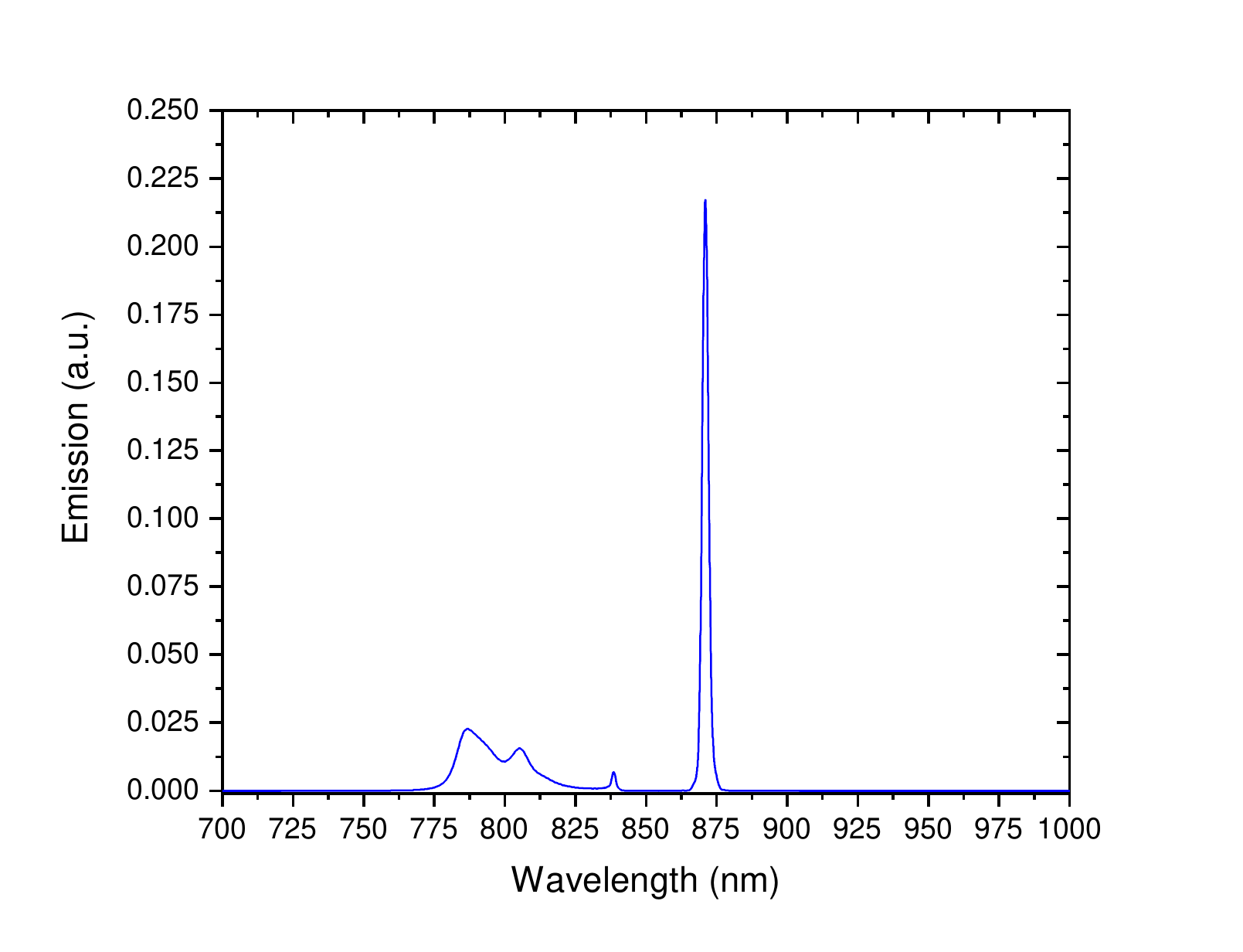}
        \captionsetup{justification=centering}
        \caption{Emission spectra in the transmission side. }    
        \label{th_red}
    \end{subfigure}
    \hfill
    \begin{subfigure}[b]{0.48\linewidth}
        \centering
        \includegraphics[width=.67\linewidth, trim= 80 30 130 90]{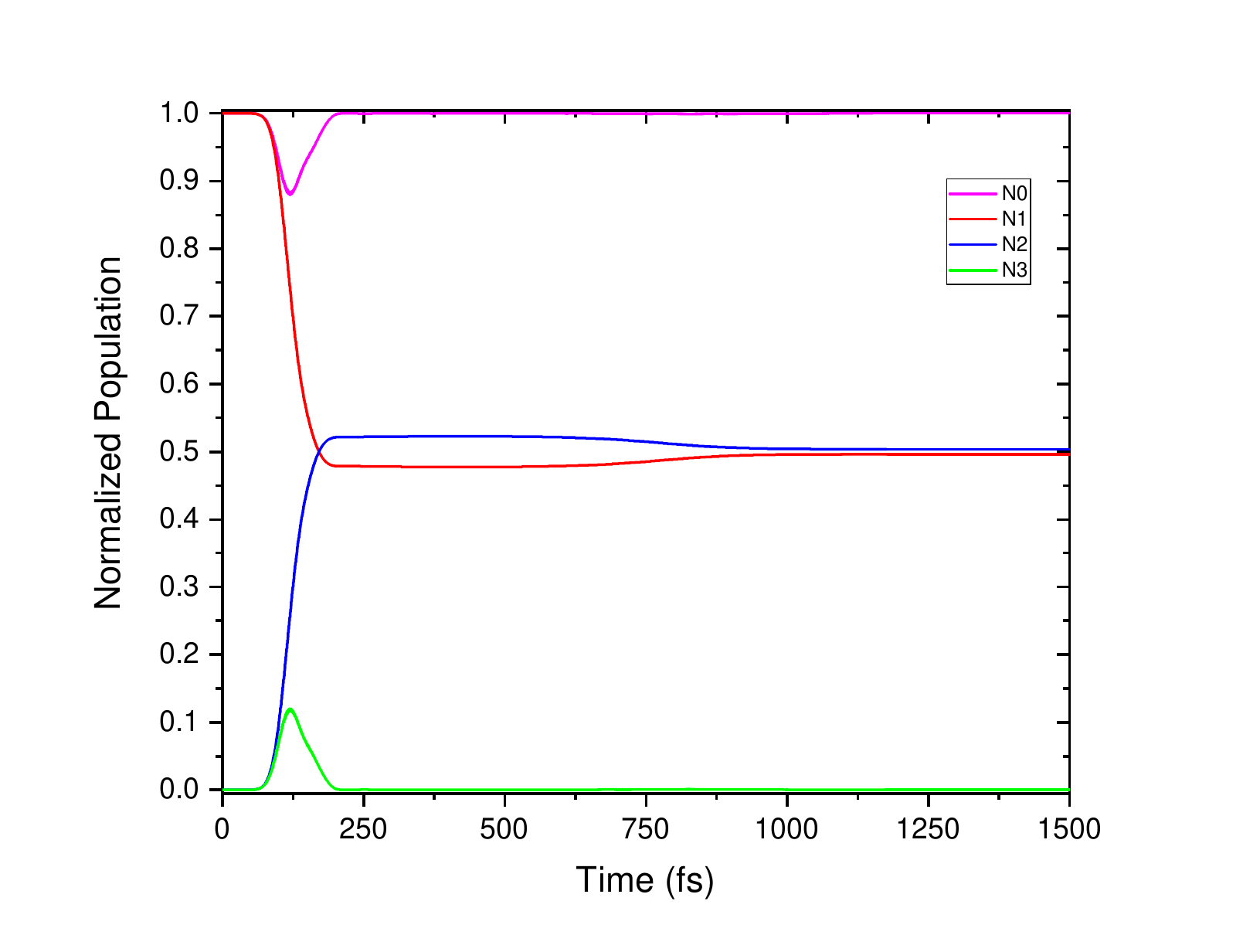}
        \caption{Population density of all four levels.}
        \label{DBR2_pop_thReduced}
    \end{subfigure}
    \captionsetup{justification=centering}
    \caption{Emission characteristics of the dual-DBR model when pumped with $2.8 \times 10^7$ V/m amplitude.\vspace{-0.4cm}}
    \label{DBR2_pop_th_red}
    \end{figure}

Although the pump amplitude in the following sections is maintained at the previously used value of $8 \times 10^7$ V/m (\(0.0256\,\text{mJ/cm}^{2}\) of pump energy) to achieve lasing at 870~nm, it is noteworthy that the required threshold pump energy has been successfully reduced by a factor of 2.857 compared to earlier reports.\cite{ahmed_efficient_2018, ahamed_wavelength_2024} The lasing emission characteristics at a pump amplitude of $2.8 \times 10^7$ V/m (\(0.0031\,\text{mJ/cm}^{2}\) of pump energy) are presented in Figure~\ref{DBR2_pop_th_red}. This result highlights a promising approach for realizing low-cost, practical nanolasers.

\vspace{-0.3cm}

\subsubsubsection{E. Far-field directionality analysis}
\vspace{-0.2cm}

In the proposed plasmonic nanolaser (PNL), lasing light couples to LSPs within the octagonal NHA holes, with each hole acting as a dipole source radiating coherent spherical waves.\cite{treshin_optical_2013} These waves interfere constructively in the far-field, producing directional emission along the surface normal. The far-field pattern is obtained via a frequency-domain field profile monitor, assuming a $70\,\mu\text{m} \times 70\,\mu\text{m}$ laser area (200 unit cells per axis) with “Top Hat” illumination.

\begin{figure}[htbp]
    \centering
    % (a) Old - Far-field distribution
    \begin{subfigure}[b]{.48\linewidth}
        \centering
        \includegraphics[width=.7\linewidth,  trim=90 230 120 240,clip]{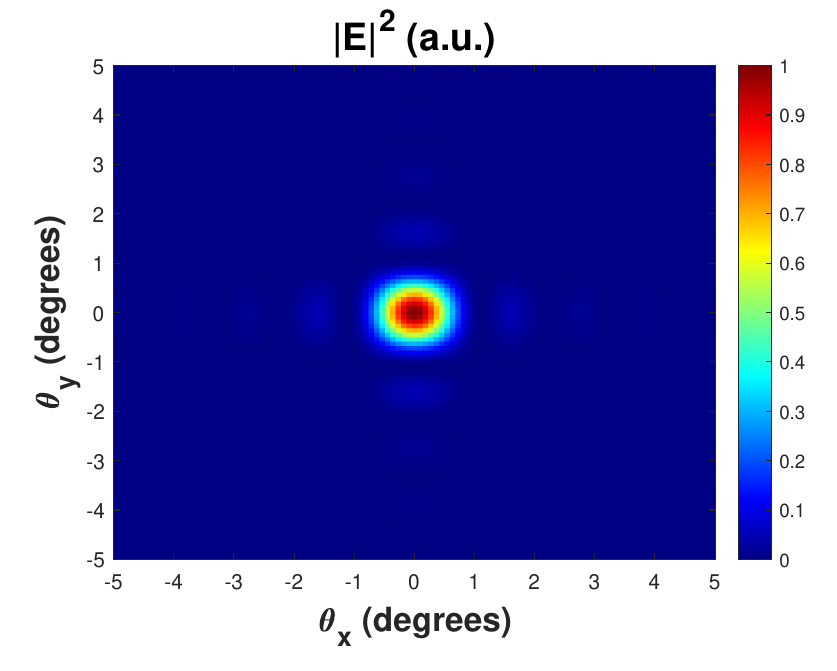}
        \caption{Far-field distribution of the PNL with top \\DBR only.}
        \label{Far_field_noDBR2}
    \end{subfigure}
    \hfill
    % (b) Old - theta_x cut
    \begin{subfigure}[b]{0.48\linewidth}
        \centering
        \includegraphics[width=.7\linewidth,  trim=90 230 120 240,clip]{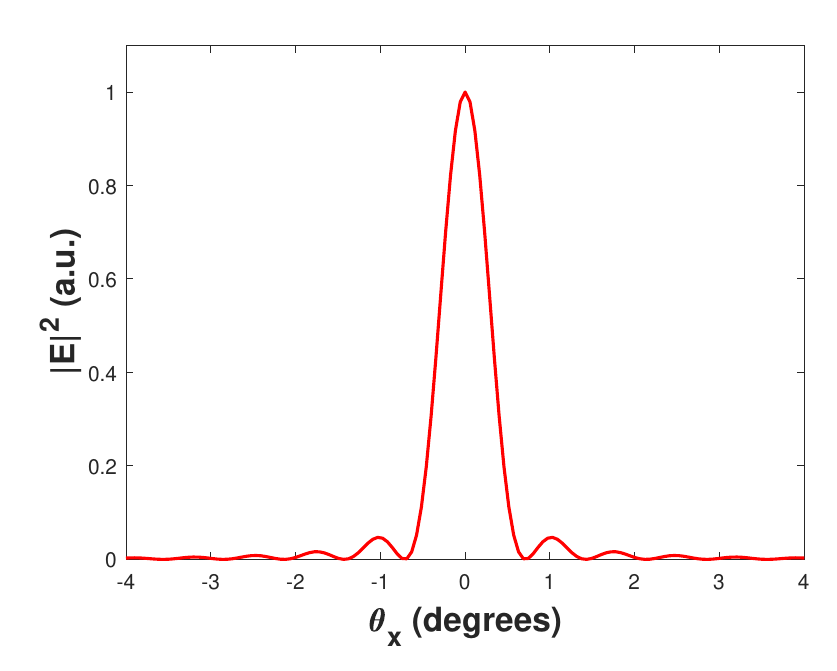}
        \caption{Far-field vs.\ $\theta_x$ ($\theta_y=0^\circ$) of the PNL with top DBR only.}
        \label{Far_field_thetax_noDBR2}
    \end{subfigure}

    % (c) New - Far-field distribution
    \begin{subfigure}[b]{.48\linewidth}
        \centering
        \includegraphics[width=.7\linewidth,  trim=90 230 120 240,clip]{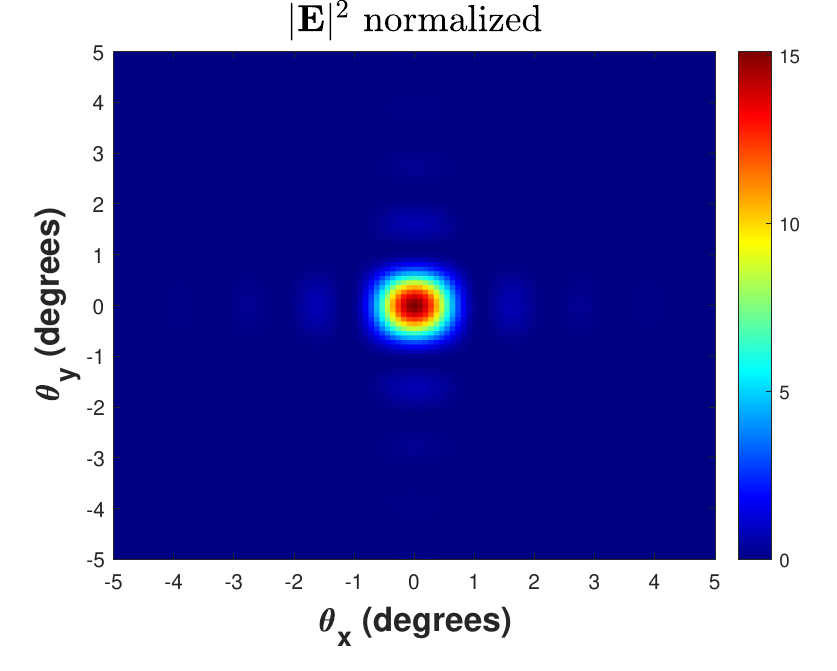}
        \caption{Far-field distribution (normalized) of the dual-DBR model.}
        \label{Far_field_DBR2}
    \end{subfigure}
    \hfill
    % (d) New - theta_x cut
    \begin{subfigure}[b]{0.48\linewidth}
        \centering
        \includegraphics[width=.7\linewidth,  trim=90 230 120 240,clip]{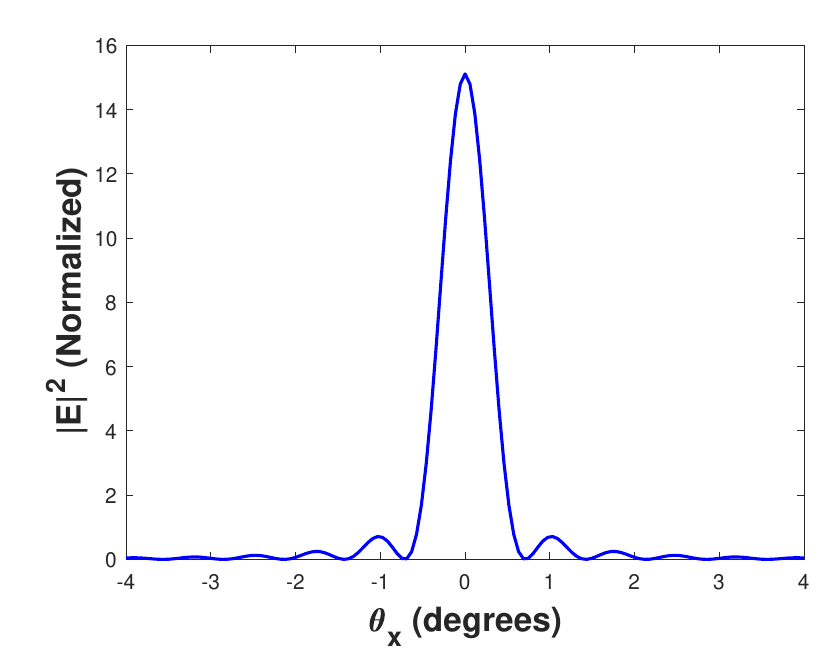}
        \caption{Far-field (Normalized) vs.\ $\theta_x$ ($\theta_y=0^\circ$) of the dual-DBR model.}
        \label{Far_field_thetax_DBR2}
    \end{subfigure}

    \caption{Far-field analysis: (a,b) previous design without DBR2 and (c,d) proposed nanolaser with bottom DBR added afterwards. Note that (c,d) are normalized using maximum intensity data obtained from (a,b) respectively.\vspace{-0.4cm}}
    \label{Far_field_Analysis_all}
\end{figure}

\begin{figure}[htbp]
    \centering
    \includegraphics[width=0.42\linewidth, trim=40 50 40 50]{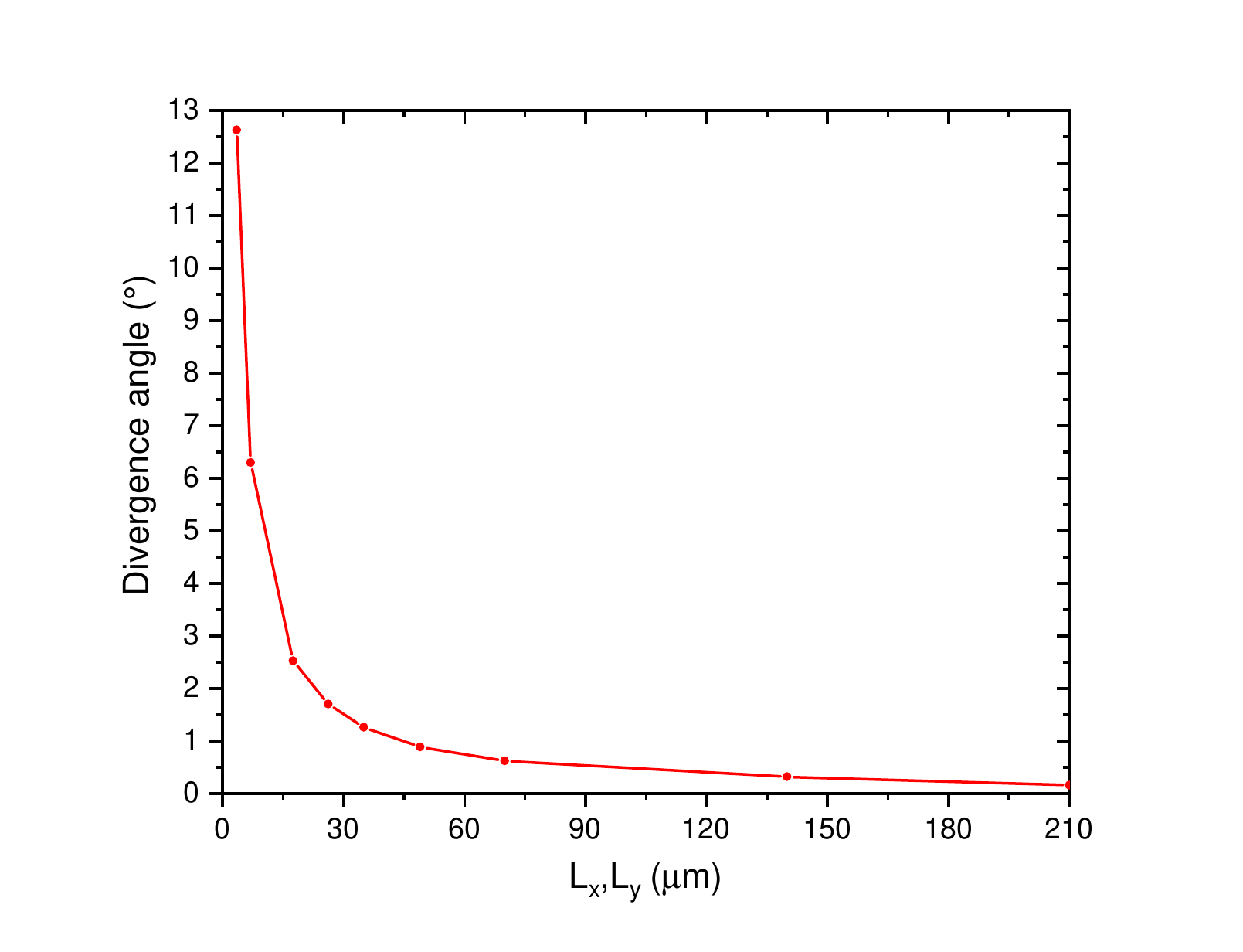}
    \caption{Divergence angle of far-field emission as the device length varies along $x$ ($L_x$) and $y$ ($L_y$) axis.\vspace{-0.4cm}}
    \label{LxLy_vary}
\end{figure}

Figure~\ref{Far_field_Analysis_all} shows the far-field intensity distribution, with $\theta_x$ and $\theta_y$ denoting angles in the $y$-$z$ and $x$-$z$ planes. Both the previous and proposed designs exhibit extremely small divergence, with FWHM $\sim 0.631^{\circ}$, outperforming Zabir et al.~\cite{ahmed_efficient_2018} who reported $\sim 1^{\circ}$ for single mode tamm lasing. Cross-sections along $\theta_x$ and $\theta_y$ (Figures~\ref{Far_field_noDBR2} and \ref{Far_field_DBR2}) show emission centered at $\theta_x=0^{\circ}$ and $\theta_y=0^{\circ}$, with the dual-DBR model reaching nearly 15 times higher peak intensity than the single-DBR design. Figures~\ref{Far_field_thetax_noDBR2} and \ref{Far_field_thetax_DBR2} highlight that in the improved model, intensity drops sharply from 15.1 a.u. to zero within $\theta_x \approx 0.68^{\circ}$, while the previous design declines more gradually, demonstrating superior directionality. Figure~\ref{LxLy_vary} shows that increasing device length along $x$ and $y$ further reduces divergence, though fabrication limits restrict practical divergence to $\sim 1^{\circ}$.\cite{meng_highly_2014}

\vspace{-0.3cm}

\subsection{\textcolor{myred}{\textbf{3.2. Tunability analysis of emission characteristics}}}
\label{Tunability}
\vspace{-0.2cm}

This section discusses the methods for tuning the lasing wavelength and emission intensity through variations in the thickness of different cavity layers, the number of layers in the top DBR, and the pump incidence angle in Figure \ref{All_tuning}. The pump amplitude is maintained at $8 \times 10^7$ V/m.

\vspace{-0.3cm}  % Adjust this value as needed

\subsubsection{3.2.1.   Tuning by the terminating and gain layer thickness}

\vspace{-0.2cm}

\begin{figure}[htbp]
    \centering

    % Subfigure 1
    \begin{subfigure}[b]{0.32\linewidth}
        \centering
        \includegraphics[width=\linewidth, trim= 60 20 50 60,clip]{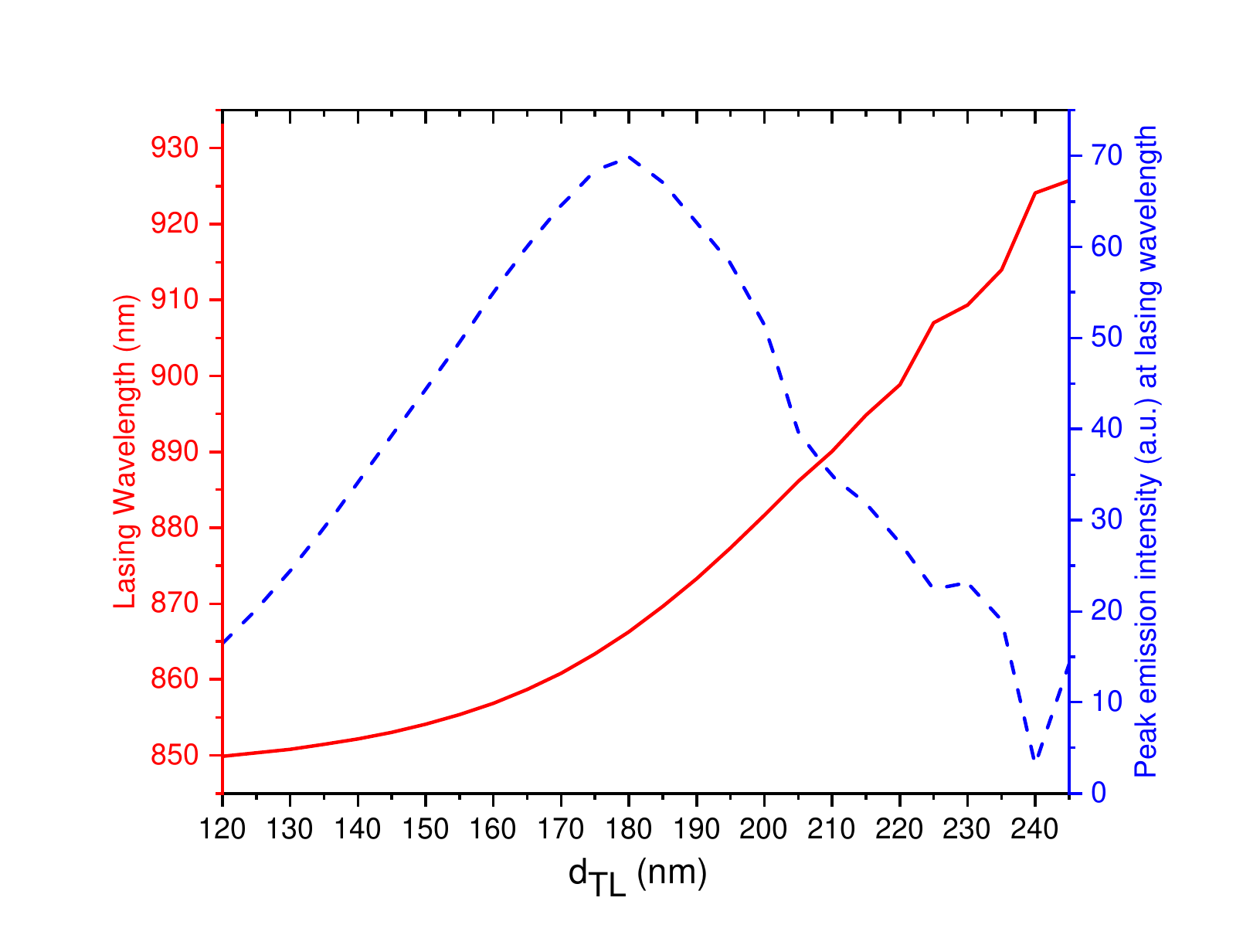}
         \captionsetup{justification=centering}
        \caption{Tuning by terminating\\ layer thickness, \(d_{\text{TL}}\).}

        \label{dtl_varied}
    \end{subfigure}
    \hfill
    % Subfigure 2
    \begin{subfigure}[b]{0.32\linewidth}
        \centering
        \includegraphics[width=\linewidth, trim= 60 20 45 60,clip]{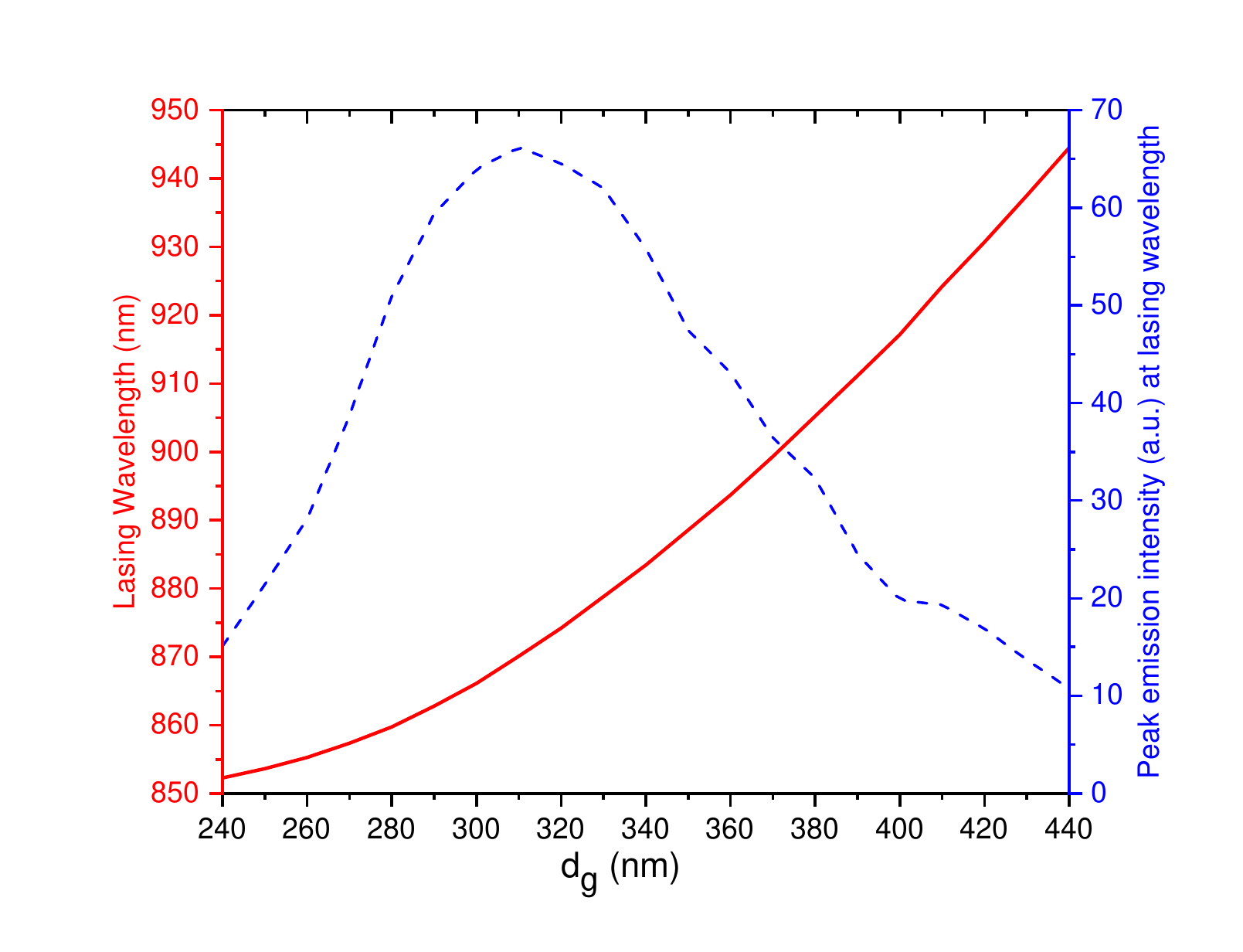}
         \captionsetup{justification=centering}
        \caption{Tuning by gain layer\\ thickness, \(d_g\).}
        \label{dg_varied}
    \end{subfigure}
    \hfill
    % Subfigure 3
    \begin{subfigure}[b]{0.32\linewidth}
        \centering
        \includegraphics[width=\linewidth, trim= 60 20 50 60,clip]{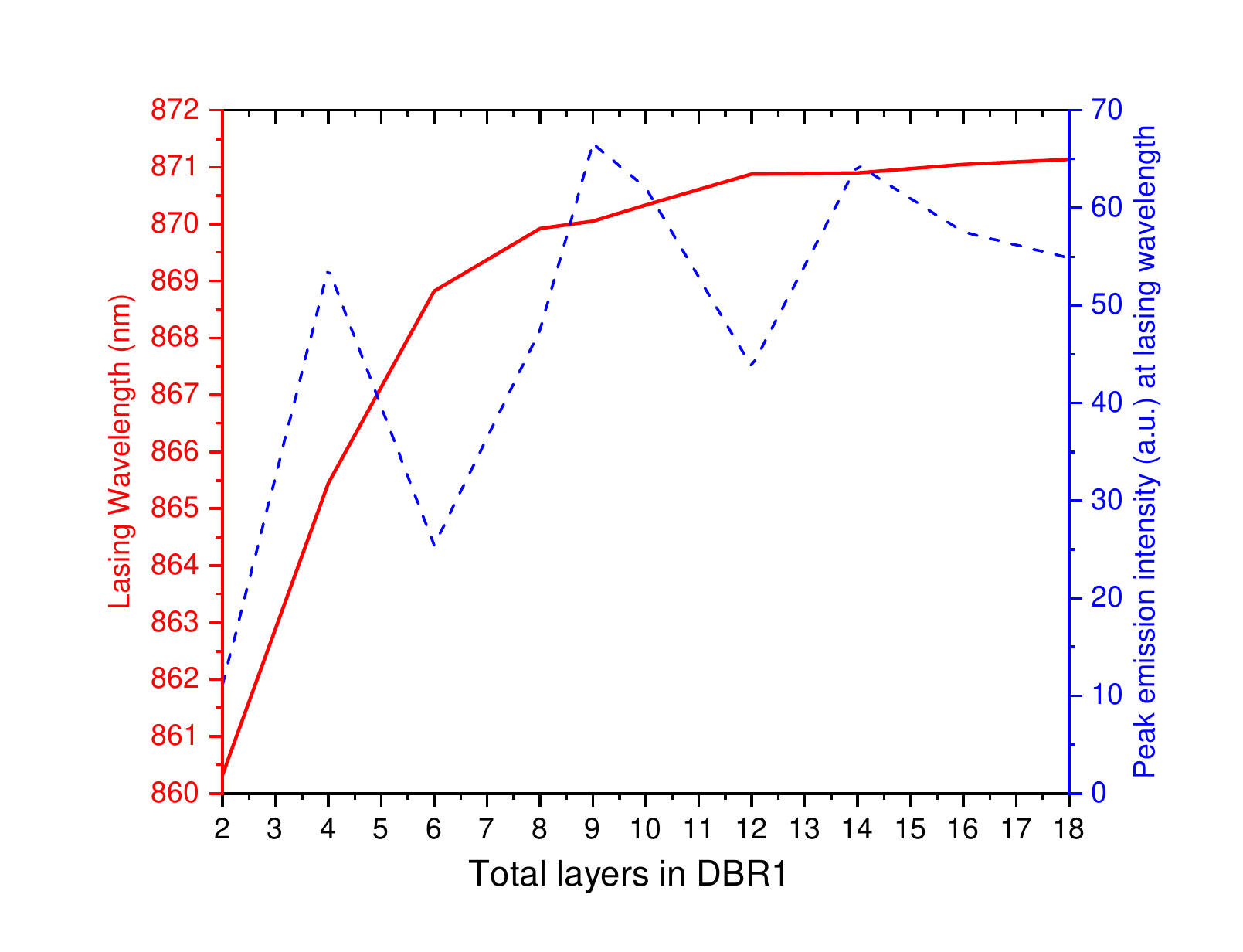}
        \captionsetup{justification=centering}
        \caption{Tuning by top DBR \\layer number.}
        \label{layers1_varied}
    \end{subfigure}
     \captionsetup{justification=centering}
    \caption{Achieving tunability in emission characteristics for the Dual-DBR PNL model.\vspace{-0.3cm}}
    \label{All_tuning}
\end{figure}

 Figure~\ref{dtl_varied} shows that as only \(d_{\text{TL}}\) is increased from 120 nm to 245 nm, the lasing wavelength shifts from approximately 850 nm to 925.725 nm. No lasing emission is observed for \(d_{\text{TL}} < 120\,\text{nm}\) or \(d_{\text{TL}} > 245\,\text{nm}\). Besides, increasing \(d_g\) from 240 nm to 440 nm enables wavelength tuning from 852.3 to 944.5 nm as shown in Figure~\ref{dg_varied}. The maximum emission of 66.2028 a.u. occurs at \(d_g = 310\,\text{nm}\) (870.055 nm). The IR-140 dye offers a broad gain centered near 870 nm, so lasing is not confined to one wavelength. The lasing wavelength depends on the cavity resonance, requiring strong feedback, confinement, and minimal loss. Changing any one layer thickness alters the DBR stopband and phase conditions, tuning the cavity resonance.\cite{shchukin_dbr_tuning, yang_tunable_plasmonic} When this resonance overlaps the dye’s gain spectrum, lasing occurs even away from 870 nm. At tuning extremes (e.g., 850 nm or 944 nm), optical gain is lower, but strong feedback and proper resonance can still sustain lasing, as shown in figures \ref{dtl_varied} and \ref{dg_varied}.

\vspace{-0.4cm}

\subsubsection{3.2.2.   Tuning by the number of layers in the top 1DPC}

\vspace{-0.2cm}

From Figure~\ref{layers1_varied}, increasing the top DBR layers (primarily responsible for the formation of the OTS) doesn't vary the lasing wavelength significantly. The emission intensity shows a non-monotonic behavior, peaking at 66.2 a.u. for 9 layers, with noticeable dips at 2, 6, and 12 layers, indicating that the optical path length becomes slightly mismatched with the cavity resonance. While too few layers fail to provide enough feedback, too many layers reduce outcoupling of the trapped photons due to high reflectivity.

\vspace{-0.35cm}

\section{4. Fabrication considerations}
\vspace{-0.18cm}

Although this study primarily focuses on numerical simulations, a potential fabrication pathway is proposed for experimental realization. A thin Al$_2$O$_3$/Ag film can be deposited on a clean substrate. The NHA regions might be patterned using lithography (e.g., electron-beam lithography), followed by Atomic Layer Etching (ALE) \cite{oxfordinstruments2025_atomiclayeretching} to form the periodic nanostructure with high precision. A PU separation layer can then be spin-coated and thermally cured. The gain medium can be formed by spin-coating a PU layer doped with IR-140 dye, with thickness control to optimize absorption near 800 nm and emission around 870 nm. On top, alternating SiO$_2$ and TiO$_2$ layers can be sequentially deposited to form the top DBR, capped with a TiO$_2$ layer. The bottom DBR can be realized by depositing alternating MgF$_2$ and TiO$_2$ layers beneath the top DBR, with thickness control to achieve high reflectivity near 870 nm. All dielectric layers can be deposited via Atomic Layer Deposition (ALD) technology.\cite{nanomaster2025_ald} Since direct vertical pumping between DBRs may be impractical, a side-illumination scheme can be employed: an external Ti:Sapphire laser (e.g., Coherent Vitara or Spectra-Physics Tsunami, 800 nm, 40 fs pulses) might be coupled laterally into the vacuum gap between the top and bottom 1DPCs using waveguiding or side-facet coupling strategies, allowing efficient excitation of the gain medium while minimizing background losses. 

Using atomic layer etching (ALE), the lateral variation of the etched octagonal nanoholes remains within $\pm 2$-$3\,\mathrm{nm}$ around the nominal diameter of $102.5\,\mathrm{nm}$. For the deposition of the $\mathrm{Al_2O_3}$ protective layer and the dielectric DBR layers, atomic layer deposition (ALD) provides angstrom-level thickness control, with a deposition precision of approximately $\pm 0.2\,\mathrm{nm}$. When the hole diameter $d_h$ varies from $99.5\,\mathrm{nm}$ to $105.5\,\mathrm{nm}$, the lasing wavelength shifts from $868.781\,\mathrm{nm}$ (with an intensity of $73.6188$ a.u.) to $868.414\,\mathrm{nm}$ (with an intensity of $55.8819$ a.u.), indicating only a minor wavelength deviation under ALE-induced dimensional variations. Similarly, varying the $\mathrm{Al_2O_3}$ thickness from $0.8\,\mathrm{nm}$ to $1.2\,\mathrm{nm}$ changes the lasing wavelength from $864.937\,\mathrm{nm}$ (58.902 a.u.) to $864.98\,\mathrm{nm}$ (59.09 a.u.), demonstrating negligible sensitivity to ALD thickness fluctuations. For the DBR layers fabricated using ALD, varying the $\mathrm{SiO_2}$ and $\mathrm{TiO_2}$ thicknesses from $169.8\,\mathrm{nm}$ and $110.8\,\mathrm{nm}$ to $170.2\,\mathrm{nm}$ and $111.2\,\mathrm{nm}$, respectively, results in lasing wavelengths ranging from $869.406\,\mathrm{nm}$ (64.75 a.u.) to $870.726\,\mathrm{nm}$ (67.35 a.u.), representing a very small deviation from the target lasing wavelength of $870\,\mathrm{nm}$. In addition, changing $d_{\mathrm{TL}}$ from $185.8\,\mathrm{nm}$ to $186.2\,\mathrm{nm}$ shifts the lasing wavelength from $869.925\,\mathrm{nm}$ (66.23 a.u.) to $870.206\,\mathrm{nm}$ (66.16 a.u.). Overall, these results from fdtd simulations indicate that the proposed fabrication technologies would introduce only negligible variations in the optical performance of the proposed dual-DBR nanolaser model, thereby enabling reliable and highly reproducible laser fabrication.

\vspace{-0.4cm}
\section{5. Conclusion}
\vspace{-0.2cm}
This work demonstrates a low-threshold, highly efficient 870 nm plasmonic nanolaser with narrow-beam emission at room temperature and broad wavelength tunability, providing a scalable pathway for on-chip photonic integration. An octagonal nanohole array in a thin silver layer with alumina couples stimulated emission from an IR-140-doped gain medium to localized surface plasmons for strong nanoscale confinement, while the top 1D photonic crystal excites optical Tamm states to enhance confinement and extraordinary optical transmission.The bottom DBR, implemented here for the first time, significantly boosts emission intensity, output power, and directionality compared to conventional single-DBR designs.Tunability via layer thicknesses and DBR number allows precise performance control. Beyond demonstrating a practical room-temperature nanolaser that overcomes DBR-related losses, this work establishes a versatile platform for future experimental studies and device optimization. The use of octagonal nanoholes for enhanced extraordinary optical transmission provides new design strategies for plasmonic cavities, while the adoption of silver instead of gold substantially reduces material costs, enabling low-cost, low-threshold devices. The dramatic improvements in output power and emission intensity achieved through the dual-DBR configuration will guide the development of next-generation plasmonic nanolasers with higher efficiency, tunability, and integration potential, ultimately supporting more sophisticated, compact, and high-performance photonic circuits for applications in on-chip communication, sensing, and quantum technologies.

\vspace{0.3cm}
\noindent\textbf{Conflicts of interest:} There are no conflicts to declare.

\vspace{0.3cm}
\noindent\textbf{Data availability:} Data for both the PNL designing and the tunability analysis are available at \\https://github.com/Tahsin0799/Tamm\_Plasmonic\_nanolaser\_NIR

\vspace{-0.35cm}

%%%%%%%%%%%%%%%%%%%%%%% References %%%%%%%%%%%%%%%%%%%%%%%%%
\printbibliography

\end{document}